\documentclass[aps]{revtex4}
\usepackage{graphicx}
\usepackage[figuresright]{rotating}

\begin{document}
\begin{flushright}
UCRL-JRNL-226480
\end{flushright}
\title{\LARGE\bf No-Core shell model for $ A = 47$ and $ A = 49$}

\author{J.~P.~Vary $^{a,b}$, A.~G.~Negoita$^{c}$, S.~Stoica$^{c}~$\\
\it $^a$
Department of Physics and Astronomy, Iowa State University, Ames, IA  50011\\
\it $^b$
Lawrence Livermore National Laboratory, Livermore, CA 94551\\
\it $^c$
Horia Hulubei National Institute for Physics and Nuclear\\
Engineering, P.O. Box MG-6, 76900 Bucharest-Magurele, Romania  }

\date{November 1, 2006}

\vspace{0.5cm}
\begin{abstract}
We apply an {\it ab-initio} approach to the nuclear structure of
odd-mass nuclei straddling $^{48}Ca$. Starting
with the NN interaction, that fits two-body scattering and bound
state data we evaluate the nuclear properties of $A = 47$ and $A =
49$ nuclei in a no-core approach.
Due to model space limitations and the absence of 3-body interactions,
we incorporate phenomenological terms determined by
fits to $A = 48$ nuclei in a previous effort.
Our modified Hamiltonian produces reasonable spectra for these odd mass
nuclei.  In addition to the differences in single-particle basis states,
the absence of a single-particle Hamiltonian in our no-core approach
obscures direct comparisons with valence effective NN interactions.
Nevertheless, we compare the fp-shell matrix elements of our initial and
modified Hamiltonians in the harmonic oscillator basis
with a recent model fp-shell interaction, the GXPF1
interaction of Honma, Otsuka, Brown and Mizusaki.
Notable differences emerge from these comparisons. In particular,
our diagonal two-body $T = 0$ matrix elements
are, on average, about 800-900keV more attractive.
Furthermore, while our initial and modified NN Hamiltonian fp-shell
matrix elements are strongly correlated, there is much less correlation
with the GXPF1 matrix elements.
\end{abstract}

\maketitle

\section{Introduction}
The low-lying levels of the A=47-49 nuclei have long been of
experimental and theoretical interest. On the one hand,
extensive experimental information about these
nuclei is available \cite{[BUR95]}-\cite{[ENDS]} and, on the other hand, this is a suitable
nuclear mass region for developing and testing effective fp-shell Hamiltonians.
Numerous detailed spectroscopic calculations have been reported.
For example, in Ref. \cite{[MZPC96]}, using
a shell model approach, Martinez-Pinedo, Zuker,
Poves and Caurier have performed full fp-shell
calculations for the A=47 and A=49 isotopes
of Ca, Sc, Ti, V, Cr and Mn.  They employed the KB3 interaction \cite{[KB3]}
with phenomenological adjustments and they
performed complete diagonalizations to obtain very
good agreement with the experimental level schemes, transition
rates and static moments.  Extensive discussions of fp-shell
effective Hamiltonians and nuclear properties can be
found in recent shell model
review articles \cite{[BAB01],[OtsukaHMSU01],[DJDean04],[Caurier05]}.

Our interest in these nuclei stems from our goal to
extend the ab-initio no-core shell model (NCSM) applications to
heavier systems than previously investigated. Until recently, the NCSM,
which treats all nucleons on an equal footing, had been limited
to nuclei up through $A=16$. However, in a recent
paper \cite{[VPSN06]} we reported the first NCSM results for
$^{48}Ca$, $^{48}Sc$ and $^{48}Ti$ isotopes, with derived and phenomenological
two-body Hamiltonians. These three nuclei are involved in double-beta decay
of $^{48}Ca$, and the interest in developing nuclear structure models for describing
them is also related to the need for accurate calculations of the nuclear
matrix elements involved in this decay. Our first goals were to see
the limitations of such an approach applied to heavier systems and
how much improvement one can obtain by adding phenomenological two-body terms
involving all nucleons.
In brief, the results were the following \cite{[VPSN06]}: i) one finds that the charge dependence
of the bulk binding energy of eight A=48 nuclei is reasonably described with an effective
Hamiltonian derived from CD-Bonn interaction\cite{Machl},
while there is an overall underbinding by about 0.4 MeV/nucleon;
ii) the resulting spectra are too compressed compared with experiment;
iii) when isospin-dependent central terms plus a tensor interaction
are added to the Hamiltonian,  one achieves accurate total binding energies for
eight A=48 nuclei and reasonable low-lying spectra for the three nuclei
involved in double-beta decay.  Only five input data were used to determine
the phenomenological terms - the total binding of $^{48}Ca$, $^{48}Sc$, and $^{48}Ti$
along with the lowest positive and negative parity excitations of $^{48}Ca$.

In the present paper we extend our previous approach to the odd-A isotopes
$^{47}Ca$, $^{49}Ca$, $^{47}Sc$ and $^{47}K$, which differ by one nucleon from $^{48}Ca$.
One of our goals is to test whether the same modified effective Hamiltonian
used for A=48 isotopes, is able to describe these odd-A nuclei.
A particular feature of the spectroscopy of these odd nuclei is that
the spin-orbit splitting gives rise to a sizable energy gap in the fp-shell
between the $f_{7/2}$ and other orbitals ($p_{1/2}$, $p_{3/2}$,
$f_{5/2}$) and we wanted to see if this feature is reproduced in the NCSM where
we have no input single-particle energies.
Also, in spite of the differences in frameworks with and without a core,
we wanted to compare our initial and modified
Hamiltonian with a recent fp-shell interaction, the GXPF1,
developed by Honma, Otsuka, Brown and Mizusaki \cite{[HOB04]}.
We feel it is valuable to compare various
fp-shell interactions in order to understand better their shortcomings and
their regimes of applicability.  From the comparison we present here, notable differences
are evident.  For example, our diagonal two-body T=0 matrix elements are more attractive and,
while our initial and modified NN Hamiltonian fp-shell matrix elements are
strongly correlated, there is much less correlation with the GXPF1 matrix elements.
It is worth mentioning that our interaction and Honma et al. GXPF1 interactions
were also tested recently within the framework of spectral distribution
theory in Ref. \cite{[SDV06]} and sizable differences were demonstrated.

Our paper is organized as follows: in Section 2 we give a short review of the NCSM
approach and we refer the reader to the bibliography for more details.
Section 3 is devoted to the presentation of our results. Binding energies, excitation spectra,
single-particle characteristics, monopole matrix elements and matrix element correlations
are discussed in subsections along with corresponding figures. In the
last Section we present the conclusion and the outlook of our work.

\section{NO-CORE SHELL MODEL}

The NCSM~\cite{ZBV,NB96,NB98,NB99,NCSM12,NCSM6} is
based on an effective Hamiltonian derived from realistic
``bare'' interactions and acting within a finite Hilbert space.
All $A$-nucleons are treated on an
equal footing.  The approach is both computationally
tractable and demonstrably convergent to the exact result
of the full (infinite) Hilbert space.

Initial investigations used two-body interactions~\cite{ZBV}
based on a G-matrix approach. Later, we implemented
a similarity transformation procedure
based on Okubo's pioneering work~\cite{Vefftot}
to derive two-body and three-body effective interactions
from realistic NN and NNN interactions.

Diagonalization and the evaluation of observables from effective
operators created with the same transformations are carried out on
high-performance parallel computers.

\subsection{Effective Hamiltonian}

For pedagogical purposes, we outline the {\it ab initio}
NCSM approach with NN interactions alone and point the reader to the
literature for the extensions to include NNN interactions.
We begin with the purely intrinsic
Hamiltonian for the $A$-nucleon system, i.e.,
\begin{equation}\label{ham}
H_A= T_{\rm rel} + {\cal V} =
\frac{1}{A}\sum_{i<j}^A \frac{(\vec{p}_i-\vec{p}_j)^2}{2m}
+ \sum_{i<j=1}^A V_{\rm N}(\vec{r}_i-\vec{r}_j) \; ,
\end{equation}
where $m$ is the nucleon mass and $V_{\rm N}(\vec{r}_i-\vec{r}_j)$,
the NN interaction, with both strong and electromagnetic components.
Note the absence of a phenomenological single-particle potential.
We may use either coordinate-space
NN potentials, such as the Argonne potentials \cite{GFMC}
or momentum-space dependent
NN potentials, such as the CD-Bonn \cite{Machl}.

Next, we add to (\ref{ham}) the center-of-mass Harmonic Oscillator (HO) Hamiltonian
$H_{\rm CM}=T_{\rm CM}+ U_{\rm CM}$, where
$U_{\rm CM}=\frac{1}{2}Am\Omega^2 \vec{R}^2$,
$\vec{R}=\frac{1}{A}\sum_{i=1}^{A}\vec{r}_i$.
At convergence, the added  $H_{\rm CM}$ term
has no influence on the intrinsic properties.
However, when we introduce our cluster approximation below,
the added  $H_{\rm CM}$ term
facilitates convergence to exact results
with increasing basis size.
The modified Hamiltonian, with pseudo-dependence on the HO
frequency $\Omega$, can be cast as:
\begin{equation}
\label{hamomega}
H_A^\Omega= H_A + H_{\rm CM}=\sum_{i=1}^A \left[ \frac{\vec{p}_i^2}{2m}
+\frac{1}{2}m\Omega^2 \vec{r}^2_i
\right]\\
+ \sum_{i<j=1}^A \left[ V_{\rm N}(ij)
-\frac{m\Omega^2}{2A}
(\vec{r}_i-\vec{r}_j)^2
\right] \; .
\end{equation}

Next, we introduce a unitary transformation, which is
designed to accommodate
the short-range two-body correlations in a nucleus,
by choosing an anti-hermitian operator $S$, acting only on
intrinsic coordinates, such that
\begin{equation}\label{UMOAtrans}
{\cal H} = e^{-S} H_A^\Omega e^{S} \; .
\end{equation}
In our approach, $S$ is determined by the requirements that ${\cal H}$
and $H_A^\Omega$ have the same symmetries and eigenspectra over
the subspace ${\cal K}$ of the full Hilbert space.
In general, both $S$ and the transformed Hamiltonian are
$A$-body operators.
Our simplest, non-trivial approximation to ${\cal H}$ is to develop
a two-body $(a=2)$ effective Hamiltonian, where the upper bound
of the summations ``$A$'' is replaced by ``$a$'', but
the coefficients remain unchanged.
We then have
an approximation at a fixed level of clustering, $a$, with $a \leq A$.
\begin{equation}\label{UMOAexpan}
{\cal H} = {\cal H}^{(1)} + {\cal H}^{(a)} =  \sum_{i=1}^{A} h_i +
\frac{{A \choose 2}}{{A \choose a}{a \choose 2}}
\sum_{i_{1}<i_{2}< \ldots <i_{a}}^{A}\tilde{V}_{i_{1}i_{2} \ldots i_{a}} \; ,
\end{equation}
with
\begin{equation}\label{UMOAexplterms}
\tilde{V}_{12 \ldots a} = e^{-S^{(a)}}H^{\Omega}_{a}e^{S^{(a)}}
- \sum_{i=1}^a h_i \; , \\
\end{equation}
and $S^{(a)}$ is an $a$-body operator; $H^{\Omega}_{a} = h_1+h_2+h_3+ \ldots
+h_{a}+V_{a}$,
and $V_{a} = \sum_{i<j}^{a} V_{ij}$.
We adopt the HO basis states that are eigenstates of
the one-body Hamiltonian $\sum_{i=1}^A h_i$.

The full Hilbert space is divided into a finite model space
(``$P$-space'') and a complementary infinite space (``$Q$-space''), using
the projectors $P$ and $Q$ with $P+Q=1$. We determine the transformation
operator $S_{a}$ from the decoupling condition
%
$Q_{a} e^{-S^{(a)}}H^{\Omega}_{a}e^{S^{(a)}} P_{a} = 0$
%
and the simultaneous restrictions $P_a S^{(a)} P_a = Q_a S^{(a)} Q_a =0$.
The $a$-nucleon-state projectors ($P_a, Q_a$)
follow from the
definitions of the $A$-nucleon projectors $P$, $Q$.

In the limit $a \rightarrow A$, we obtain the exact
solutions for $d_P$ states of the full problem for any finite basis space, with
flexibility for choice of physical states subject to certain
conditions~\cite{Viaz01}. This approach has a significant residual freedom
through an
arbitrary residual $P_a$--space unitary transformation that leaves
the $a$-cluster properties invariant.   Of course, the $A$-body results
obtained with the a-body cluster approximation are not
invariant under this residual transformation.  An effort is underway to exploit
this residual freedom to accelerate convergence in practical
applications.

The model space, $P_2$, is defined by $N_{\rm m}$ via the maximal
number of allowed HO quanta of the $A$-nucleon basis states, $N_{\rm
M}$, where the sum of the nucleons'
$2n+l\leq N_{\rm m}+N_{\rm spsmin}=N_{\rm M}$, and where $N_{\rm
spsmin}$ denotes the minimal possible HO quanta of the spectators, 
nucleons not involved in the interaction.
For example, $^{10}$B,
$N_{\rm spsmin}=4$ as there are 6 nucleons in the $0p$-shell
in the lowest HO configuration and, e.g.,
$N_{\rm m}=2+N_{\rm max}$, where $N_{\rm max}$
represents the maximum HO quanta of the many-body excitation
above the unperturbed ground-state configuration.
For $^{10}$B, $N_{\rm M}=12, N_{\rm m}=8$
for an $N_{\rm max}=6$ or ``$6\hbar\Omega$'' calculation.
With our cluster approximation,
a dependence of our results on $N_{\rm max}$
(or equivalently, on $N_{\rm m}$ or $N_{\rm M}$) and on
$\Omega$ arises.
The residual $N_{\rm max}$ and $\Omega$ dependences
will infer the uncertainty in our results
arising from effects associated with increasing $a$ and/or
effects with increasing $N_{\rm max}$.  In the present work, we
retain the $N_{\rm max}=0$ basis space and $\hbar\Omega=10 MeV$
employed in Ref. \cite{[VPSN06]}.

At this stage
we also add the term $H_{CM}$ again with a large positive coefficient
(constrained via Lagrange multiplier)
to separate the physically interesting states with $0s$ CM motion from
those with excited CM motion. We diagonalize the effective Hamiltonian
with the m-scheme Lanczos method to obtain the $P$-space
eigenvalues and eigenvectors \cite{Vary_code}.
All observables are then evaluated free of CM motion effects \cite{Vary_code}.
In principle, all observables require the same transformation as
implemented for the
Hamiltonian.  We obtain small renormalization effects on long range operators
such as the rms radius operator and the $B(E2)$ operator
when we transform them
to $P$-space effective operators at the $a=2$ cluster
level~\cite{NCSM12,Stetcu}.
On the other hand, when a=2, substantial renormalization was observed for the kinetic
energy operator ~\cite{benchmark01}.
and for higher momentum transfer observables~\cite{Stetcu}.

Recent applications include:
\begin{itemize}
\item[(a)] spectra and transition rates in $p$-shell nuclei \cite{Nogga06};
\item[(b)] comparisons between NCSM and Hartree-Fock \cite{Hasan03};
\item[(c)] di-neutron correlations in the $^6$He halo
   nucleus \cite{Atramentov05};
\item[(d)] neutrino cross sections on $^{12}$C \cite{HayesPRL};
\item[(e)] novel NN interactions using inverse scattering theory plus NCSM \cite{Shirokov04, Shirokov05,Shirokov06};
\item[(f)] plus others in nuclear theory and quantum field theory\cite{Vary06}.
\end{itemize}

We close this theory overview by referring to the added phenomenological NN interaction terms found
adequate for obtaining good descriptions of $A=48$ nuclei \cite{[VPSN06]}.  Three terms are added -
central modified gaussians with isospin dependent strengths and a tensor term.  NCSM results obtained below with the modified Hamiltonian are referred to with "CD-Bonn + 3 terms" results. It is our hope that
these terms accommodate, to a large extent, the missing many-body forces, both real and effective.  This hope will be tested in the future when increasing computational resources allow larger basis spaces, improved $a=3$ and $a=4$ calculations as well as the introduction of true NNN and NNNN potentials.

\section{RESULTS AND DISCUSSION}

\subsection{Binding energies}

First we present the calculated total interaction energies
(Hamiltonian ground state eigenvalues)
in Fig. 1 which we compare with experiment.
One observes that ground states calculated with our derived $\it{ab-initio}$
$H_{eff}$ lie above the experimental values by approximately 20MeV.
This shift is similar to that
observed in the case of all $A=48$ isotopes \cite{[VPSN06]}.
We note that with CD-Bonn we have nearly the same increase in binding from
$^{47}Ca$ to $^{48}Ca$ as from $^{48}Ca$ to $^{49}Ca$,
which signals a lack of  subshell closure.

For the modified Hamiltonian (CD-Bonn + 3 terms) the NCSM produces
reasonable agreement with experiment with deviations much less than 1\%
as seen in Fig. 1.  There is a simple spreading of the
theoretical ground states relative to experiment.
In particular,  we now observe the desired subshell closure condition
where the increased binding from $^{47}Ca$ to $^{48}Ca$
significantly exceeds that from  $^{48}Ca$ to $^{49}Ca$.

\begin{figure}[htb]
\begin{minipage}[t]{80mm}
\framebox[79mm]{\rule[0mm]{0mm}{52mm}\includegraphics[scale=0.32]{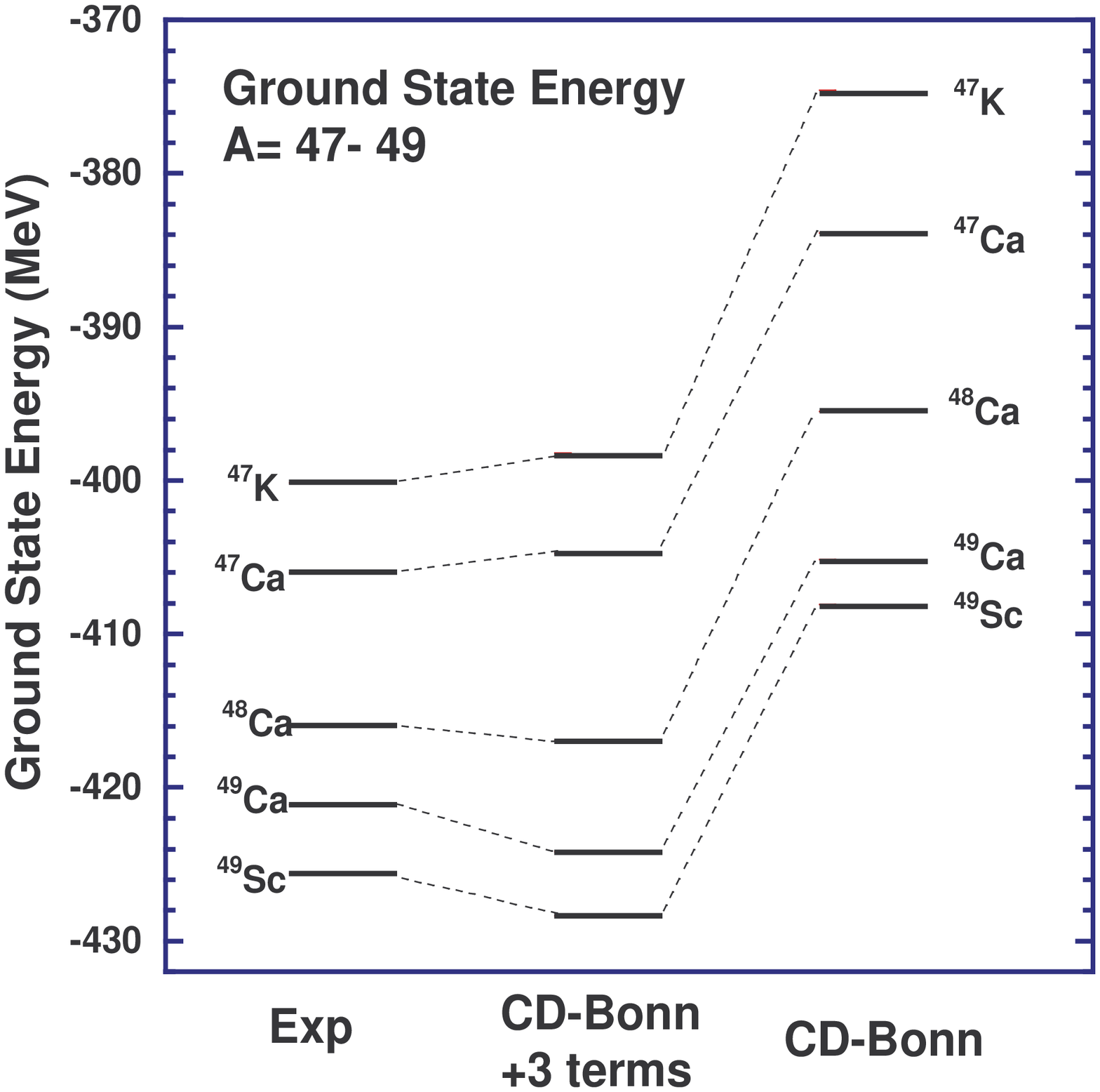}}\caption{The experimental and theoretical ground state energy levels
for A=47-49. The results in the second and third columns are labelled by
their Hamiltonians.}
\label{fig:Ground-state-energy}
\end{minipage}
\hspace{\fill}
\begin{minipage}[t]{84mm}
\framebox[80mm]{\rule[0mm]{0mm}{52mm}\includegraphics[scale=0.38]
{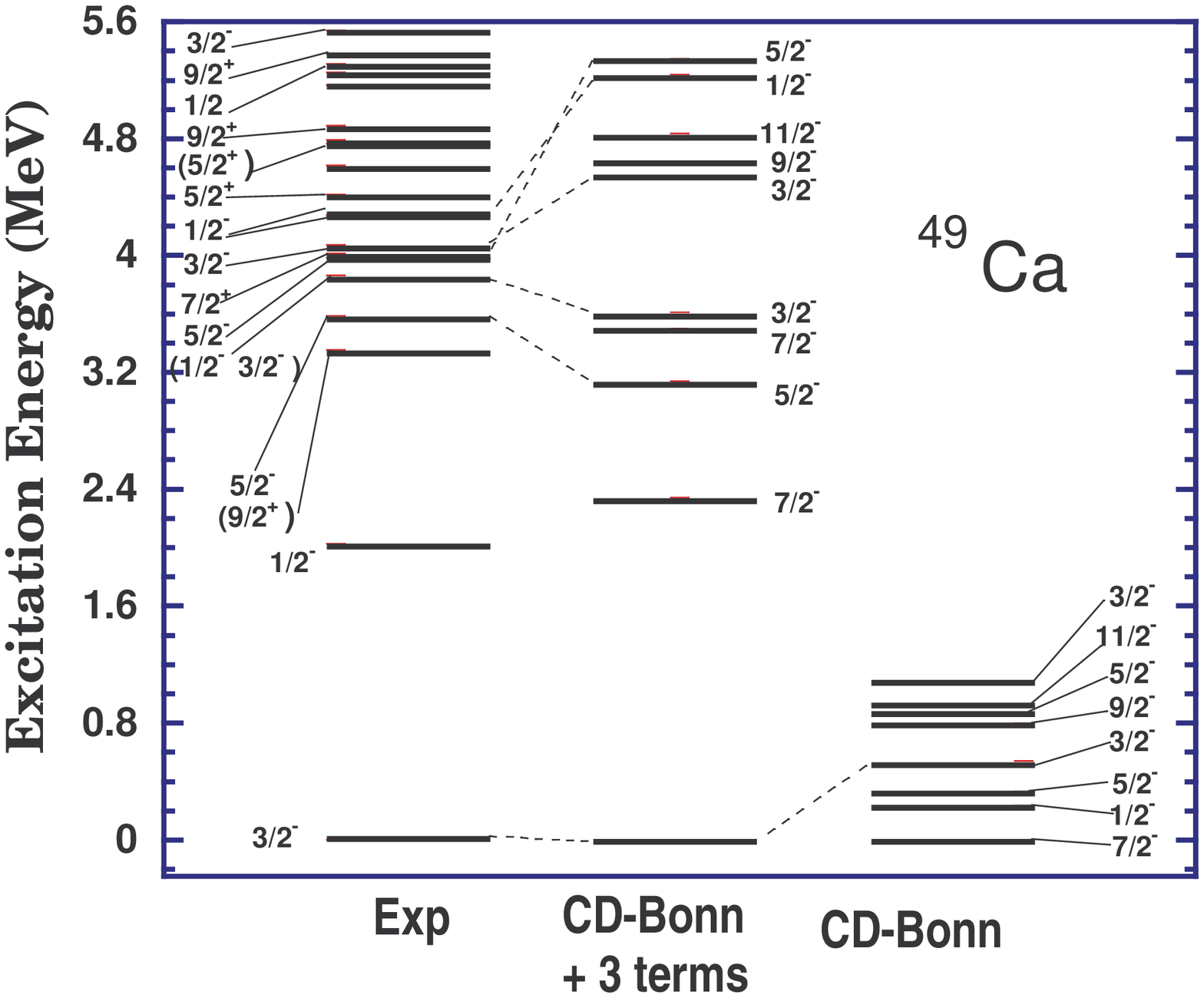}}
\caption{Experimental and theoretical excitation energy levels for
$^{49}{Ca}$. Both CD-Bonn and CD-Bonn+3terms resultss are presented.}
\label{fig:Ca49}
\end{minipage}
\end{figure}

\begin{figure}[htb]
\begin{minipage}[t]{75mm}
\framebox[74mm]{\rule[0mm]{0mm}{52mm}\includegraphics[scale=0.32]
{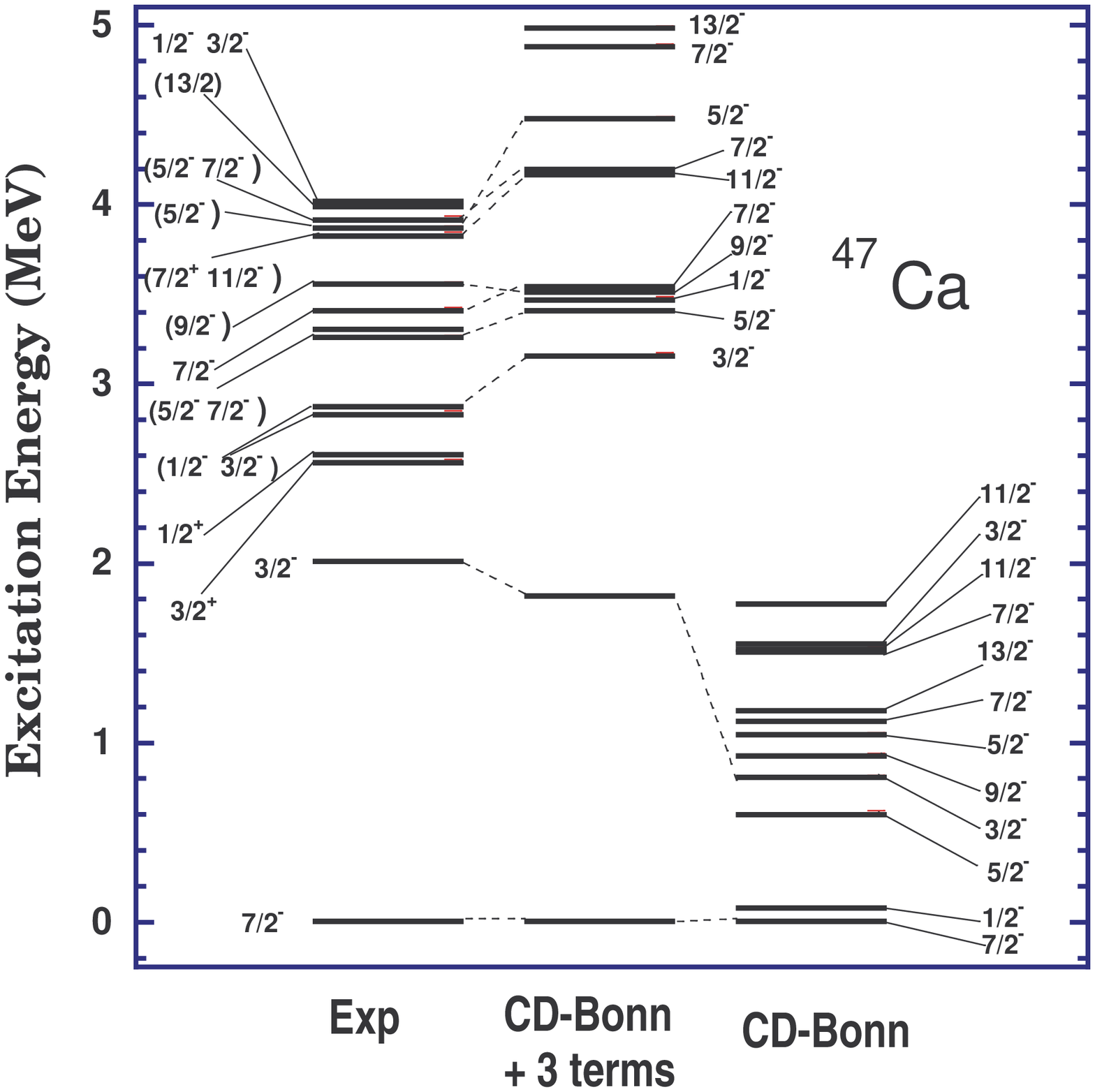}}
\caption{Experimental and theoretical excitation energy levels for
$^{47}{Ca}$. Both CD-Bonn and CD-Bonn+3terms results are presented.}
\label{fig:Ca47}
\end{minipage}
\hspace{\fill}
\begin{minipage}[t]{75mm}
\framebox[79mm]{\rule[0mm]{0mm}{52mm}\includegraphics[scale=0.32]
{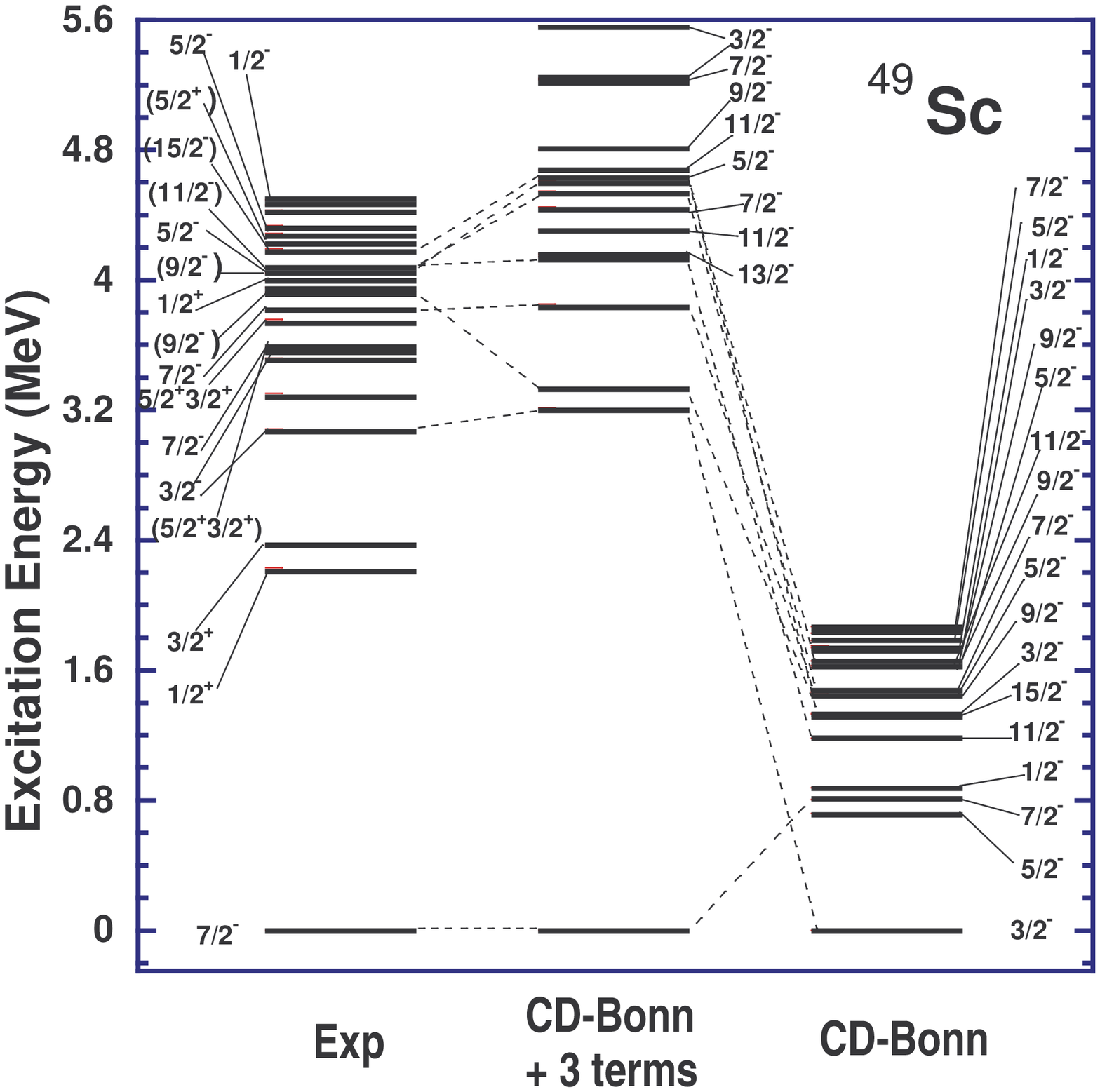}}
\caption{Experimental and theoretical excitation energy levels for
$^{49}{Sc}$. Both CD-Bonn and CD-Bonn+3terms results are presented.}
\label{fig:Sc49}
\end{minipage}
\end{figure}

\begin{figure}[htb]
\begin{minipage}[t]{75mm}
\framebox[74mm]{\rule[0mm]{0mm}{52mm}\includegraphics[scale=0.32]
{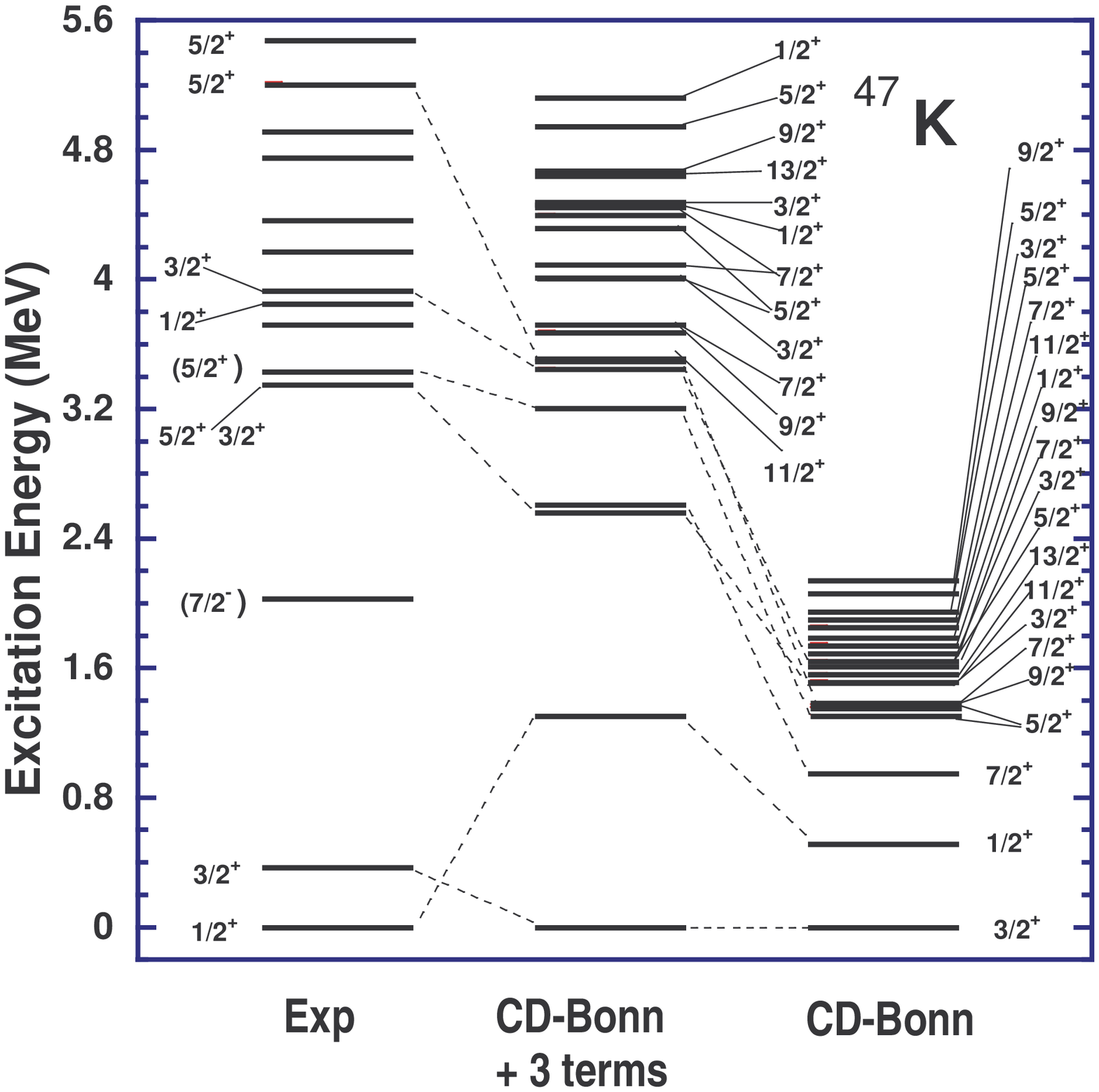}}
\caption{Experimental and theoretical excitation energy levels for
$^{47}{K}$. Both CD-Bonn and CD-Bonn+3terms results are presented.}
\label{fig:K47}
\end{minipage}
\hspace{\fill}
%
\begin{minipage}[t]{75mm}
\framebox[74mm]{\rule[0mm]{0mm}{52mm}\includegraphics[scale=0.32]
{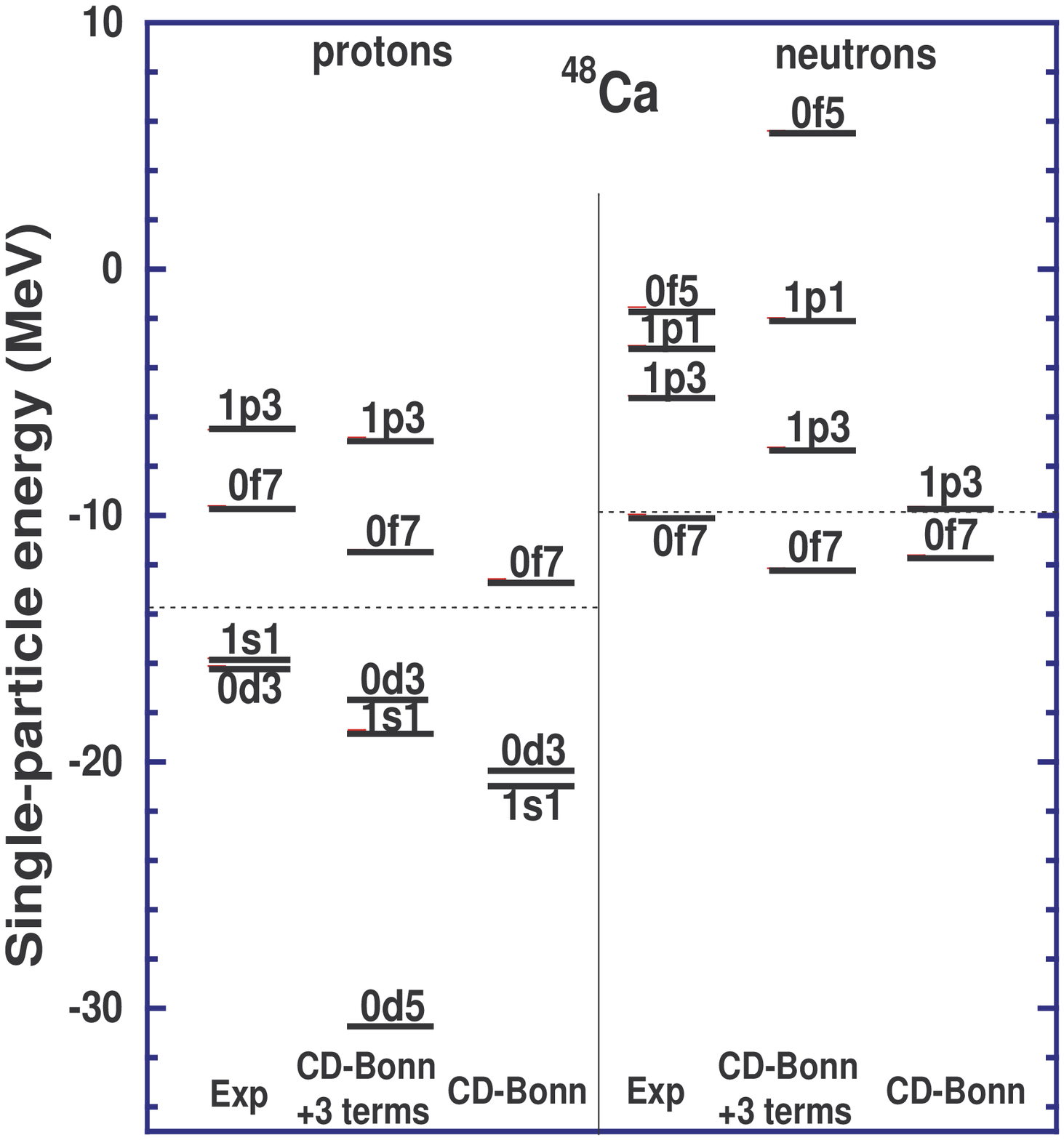}}
\caption{Experimental and theoretical levels that are dominantly
single-proton and single-neutron particles or holes of $^{48}{Ca}$.
The levels are labeled by  (n, l, 2j), and the dashed lines are the
Fermi energies.}
\label{fig:Single-particle-energy}
\end{minipage}
\end{figure}

\begin{figure}[htb]
\begin{minipage}[t]{80mm}
\framebox[79mm]{\rule[0mm]{0mm}{52mm}\includegraphics[scale=0.32]{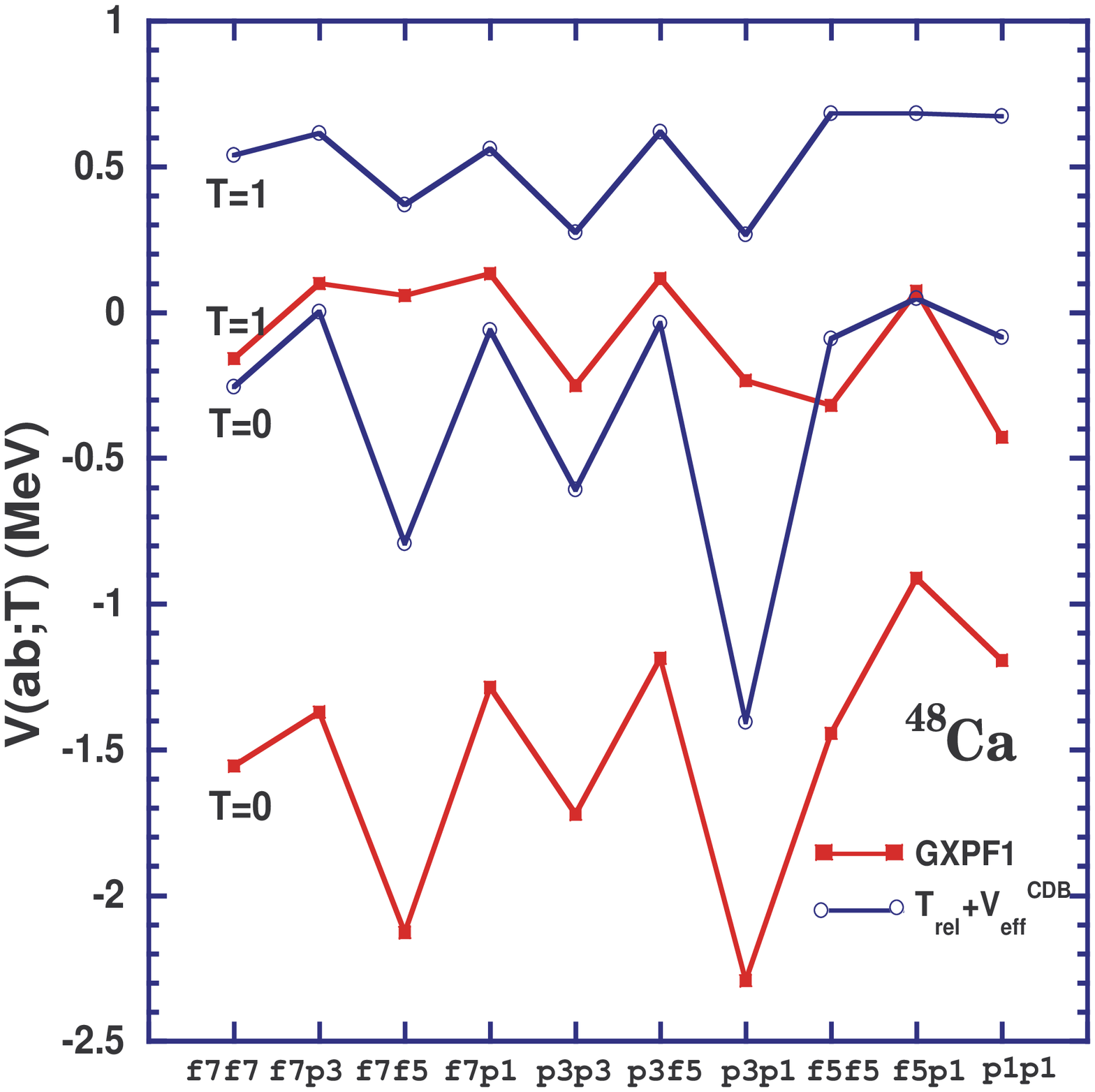}}
\caption{(color online) Comparison between CD-Bonn and GXPF1 of the
monopole matrix elements V(ab;T)(A=48), which are shown by circles and squares,
respectively. Lines are drawn to guide the eyes.
The orbit pair label "f7p3" stands for $a=f_{7/2}$ and $b=p_{3/2}$,
for example.}
\label{fig:CD-Bonn}
\end{minipage}
\hspace{\fill}
\begin{minipage}[t]{75mm}
\framebox[74mm]{\rule[0mm]{0mm}{52mm}\includegraphics[scale=0.32]
{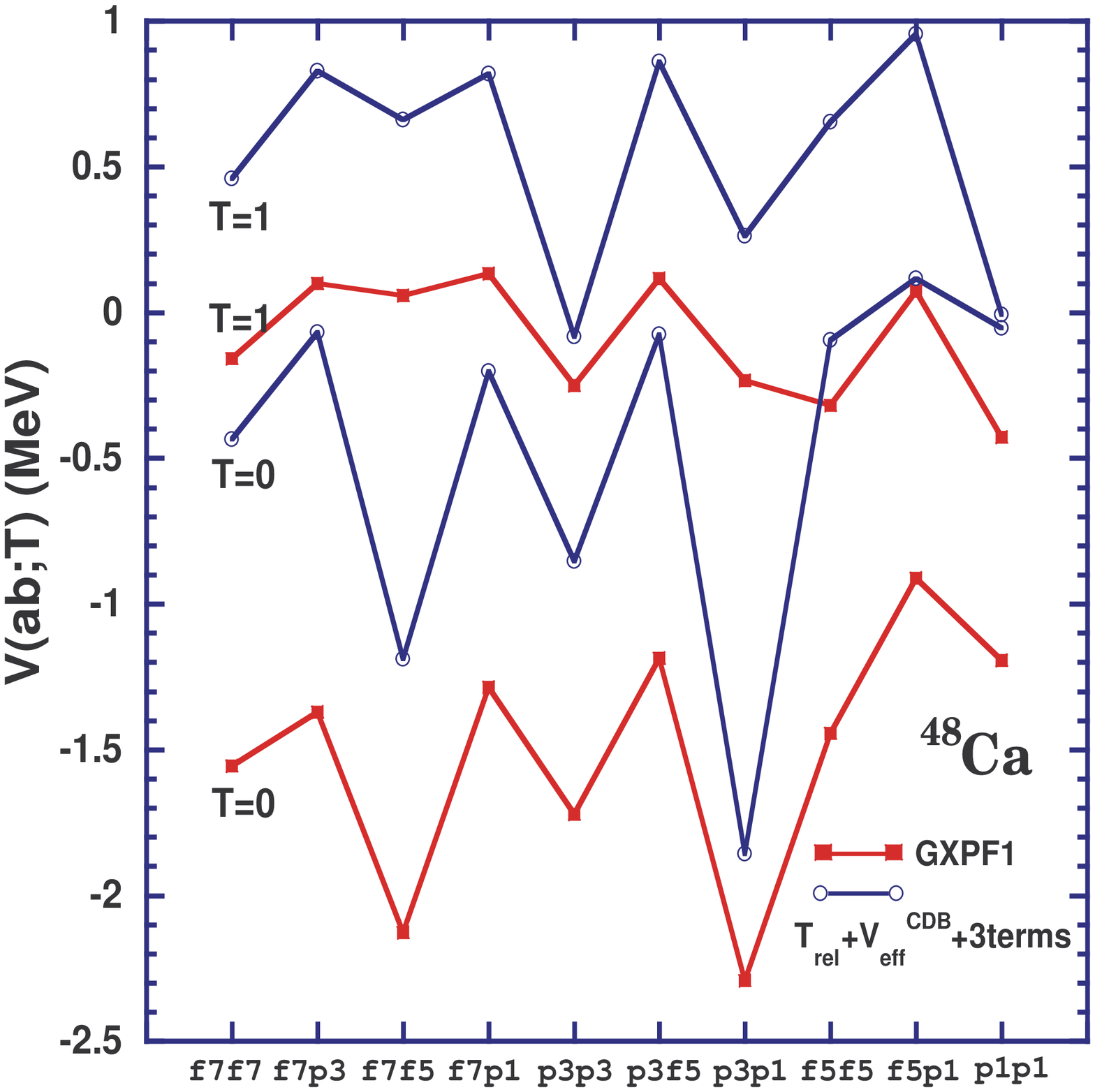}}
\caption{(color online)Comparison between CD-Bonn+3terms and GXPF1 of the
monopole matrix elements V(ab;T)(A=48), which are shown by circles and squares,
respectively. See the caption to Fig. 7.}
\label{fig:CDB+3terms}
\end{minipage}
\end{figure}

\begin{figure}[htb]
\begin{minipage}[t]{80mm}
\framebox[79mm]{\rule[0mm]{0mm}{52mm}\includegraphics[scale=0.32]
{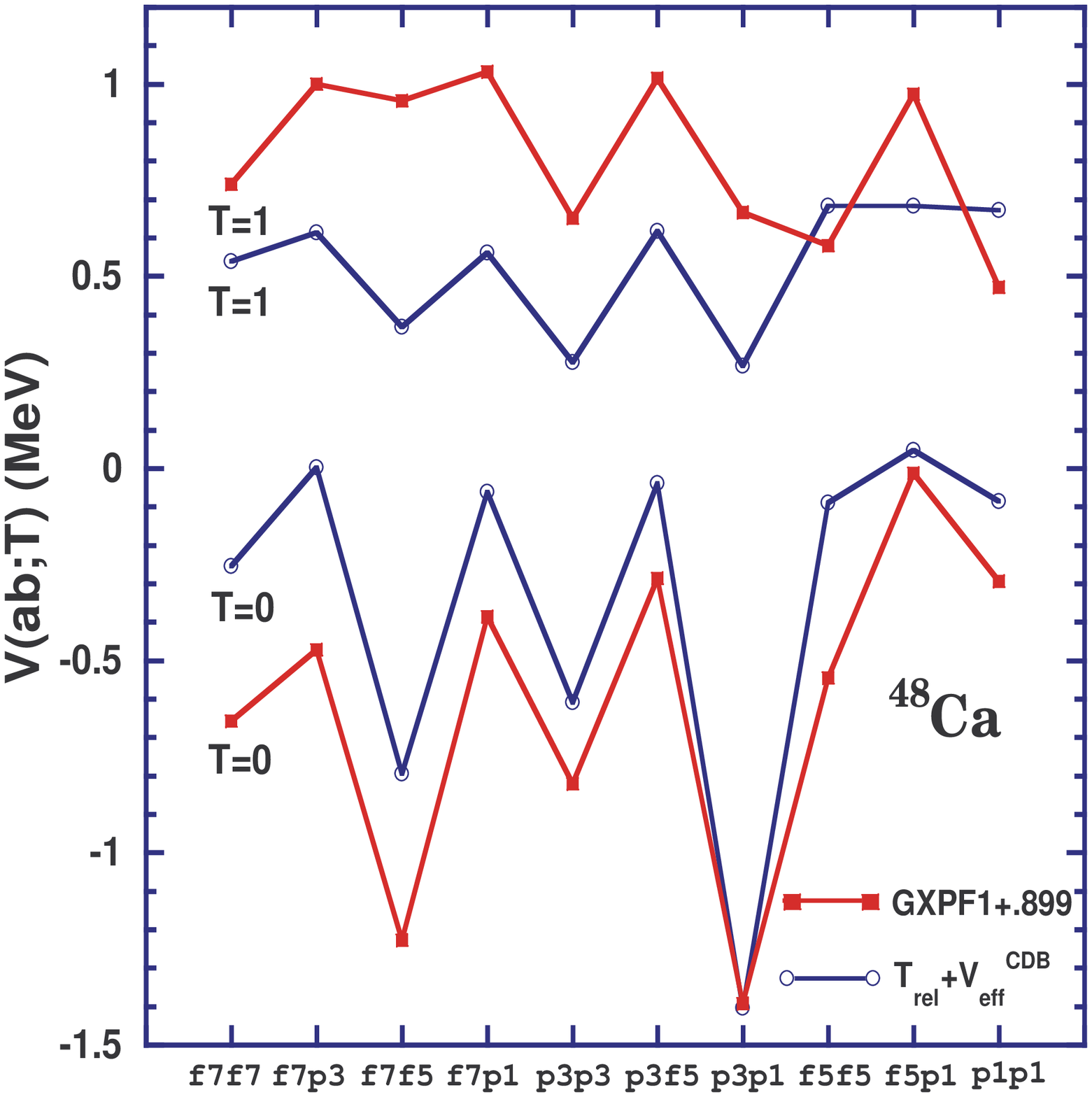}}
\caption{(color online) Comparison of the monopole matrix elements V(ab;T)(A=48)
between CD-Bonn and GXPF1 (shifted to have an overall average monopole
the same as CD-Bonn), which
are shown by circles and squares, respectively. See the caption to Fig. 7.}
\label{fig:Shift-CD-Bonn}
\end{minipage}
\hspace{\fill}
\begin{minipage}[t]{80mm}
\framebox[79mm]{\rule[0mm]{0mm}{52mm}\includegraphics[scale=0.32]
{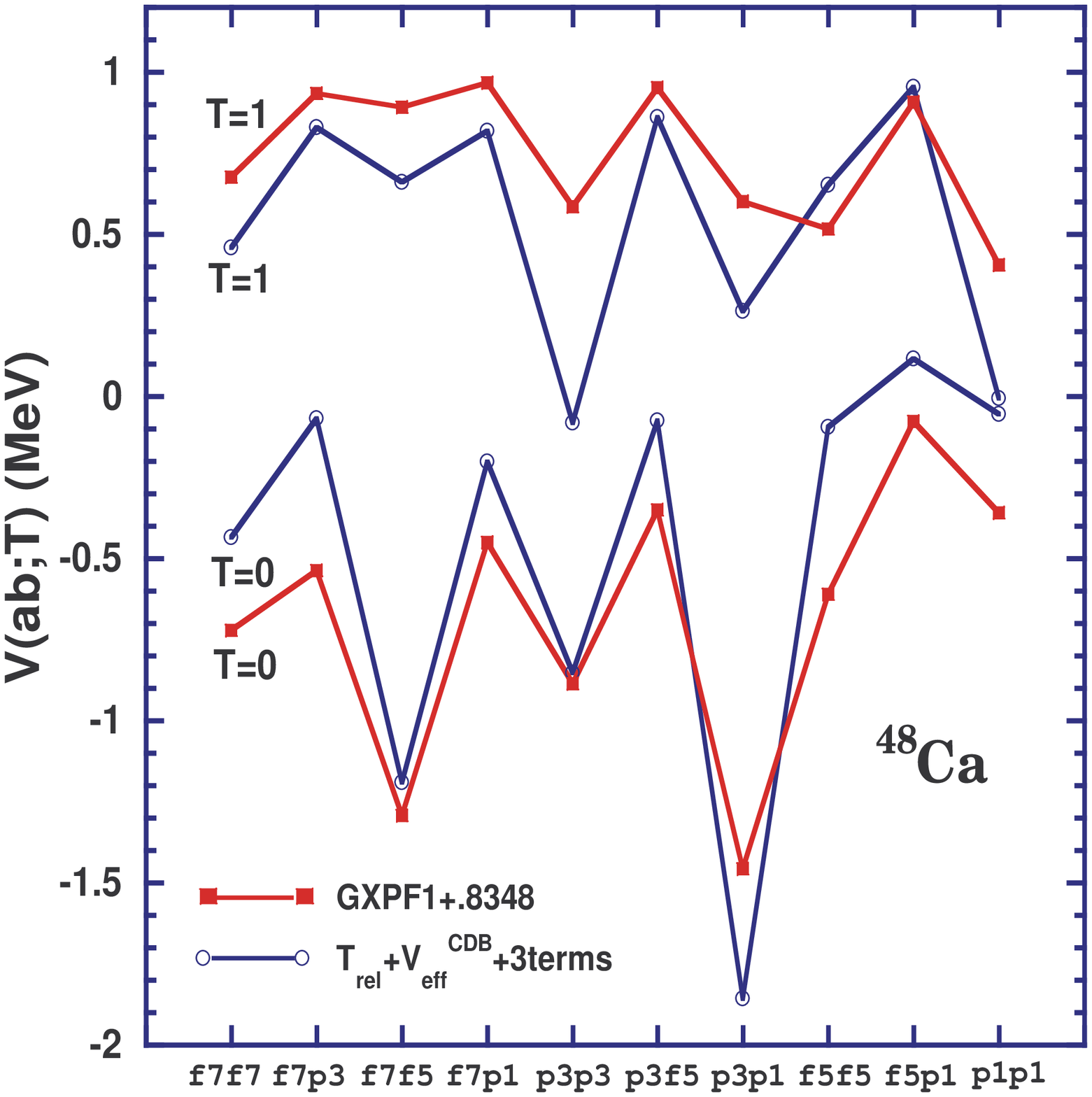}}
\caption{(color online) Comparison of the monopole matrix elements V(ab;T)(A=48) between CD-Bonn+3terms
and GXPF1 (shifted to have an overall average monopole
the same as CD-Bonn + 3 terms), which are shown
by circles and squares, respectively. See the caption to Fig. 7.}
\label{fig:Shift-CD-Bonn+3terms}
\end{minipage}
\end{figure}

\begin{figure}[htb]
\begin{minipage}[t]{80mm}
\framebox[79mm]{\rule[0mm]{0mm}{52mm}\includegraphics[scale=0.32]
{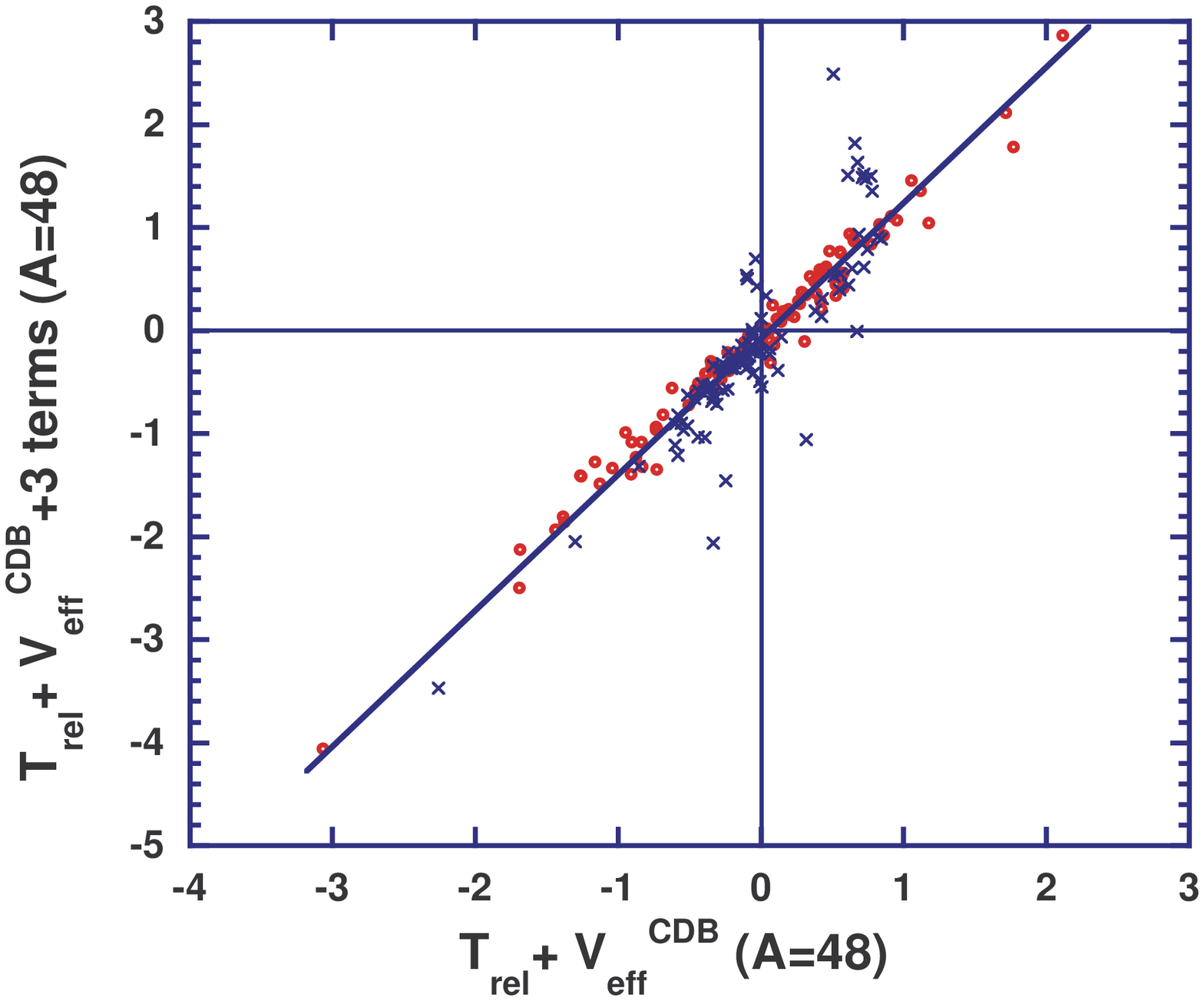}}
\caption{(color online) Correlation of V(abcd; JT) (A=48) matrix elements between
CD-Bonn+3terms and CD-Bonn.
The matrix elements of T=0 and T=1 are shown 
by open circles and crosses, respectively. There are no monopole shifts.
The straight line represents a linear fit to all the matrix elements.}
\label{fig:Fig11(Monopole)}
\end{minipage}
\hspace{\fill}
\begin{minipage}[t]{75mm}
\framebox[74mm]{\rule[0mm]{0mm}{52mm}\includegraphics[scale=0.32]
{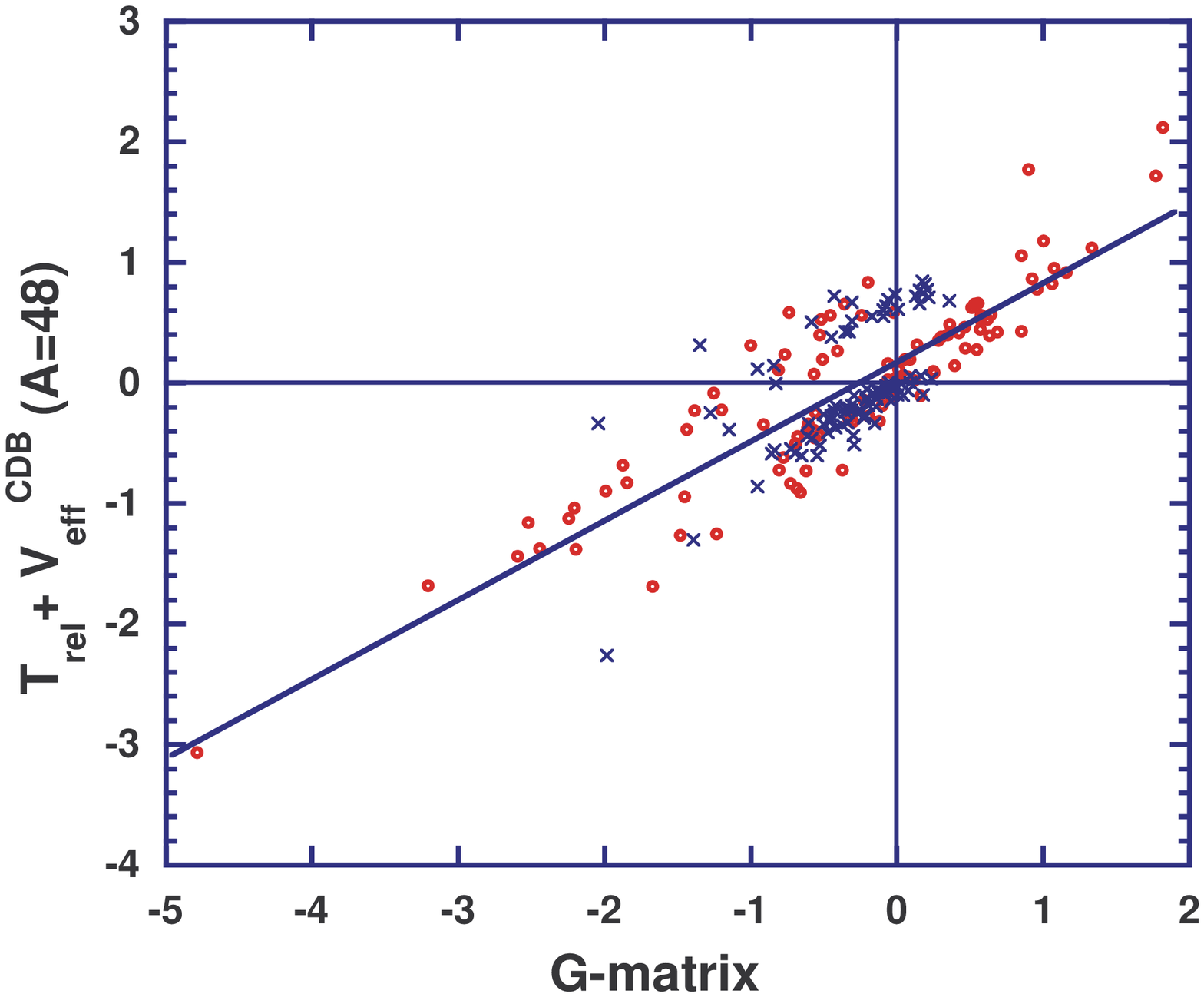}}
\caption{(color online) Correlation of V(abcd; JT)(A=48) between CD-Bonn and G. 
See the caption to Fig. 11.}
\label{fig:CD-Bonn(G)1}
\end{minipage}
\end{figure}

\begin{figure}[htb]
\begin{minipage}[t]{75mm}
\framebox[74mm]{\rule[0mm]{0mm}{52mm}\includegraphics[scale=0.32]
{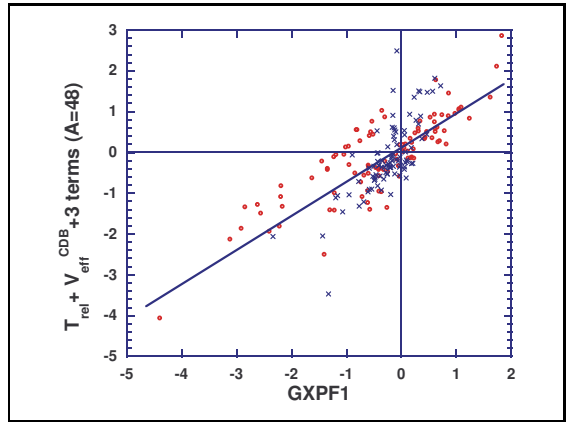}}
\caption{(color online) Correlation of the matrix elements V(abcd; JT)(A=48) between
CD-Bonn+3terms and GXPF1.
See the caption to Fig. 11.}
\label{fig:Fig.22(Monopole)}
\end{minipage}
\end{figure}

\begin{figure}[htb]
\begin{minipage}[t]{75mm}
\framebox[74mm]{\rule[0mm]{0mm}{52mm}\includegraphics[scale=0.32]
{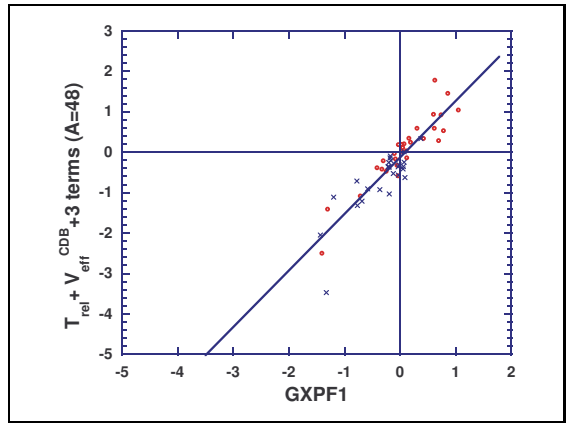}}
\caption{(color online) Correlation of the matrix elements V(abcd; JT)(A=48) between
CD-Bonn+3terms and GXPF1,
where we retain only the off-diagonal matrix elements, i.e. the 56 matrix
elements that cannot contribute to a single-particle Hamiltonian (see text).
See the caption to Fig. 11.}
\label{fig:True Off-Diag}
\end{minipage}
\end{figure}

\subsection{Excitation energy spectra}

\noindent
The excitation energy spectra for $^{49}Ca$, $^{47}Ca$,
$^{49}Sc$ and $^{47}K$ are shown in Figs. 2-5 respectively.
In every case the $\it{ab-initio}$ NCSM results with CD-Bonn
are far too compressed relative to experiment - a feature also seen
in the $A=48$ results \cite{[VPSN06]}.  Here, we trace this
primary defect to the inferred properties of the neutron orbits.
That is, the incorrect ground state
spin seen in Fig. 2 and the absence of a significant excitation energy
gap in Fig. 3 indicate the spin-orbit splitting of the neutrons is
insufficient to provide proper subshell closure at the neutron
$0f_{7/2}$ orbit.  This defect is rectified in the results
with CD-Bonn + 3 terms as seen by the corresponding spectra
in Figs. 2 and 3. Similar tendencies have been seen before
with valence G-matrix interactions and identified as a problem with
the $L^{2}$ dependence of the single-particle states
\cite{[MZPC96],[Caurier05]}.

The CD-Bonn results in Figs. 4 and 5 are more difficult to interpret due
to the glaring deficiencies just mentioned for the neutrons with the CD-Bonn
Hamiltonian.  We will show below that the proton shell closure
is better established with CD-Bonn.  This supports the assertion that
the main deficiencies seen in the third columns of Figs. 4 and 5 are
indeed likely to reside with the inferred neutron spin-orbit splitting
problem.

The modified Hamiltonian provides greatly improved spectra for
all four nuclei as seen in the second columns of Figs. 2-5.
It is to be noted that these nuclei were not involved
in the fitting procedure used to determine the parameters of the added
phenomenological terms.  Perhaps the most significant remaining
deficiency is the incorrect ground state spin for $^{47}K$ as seen in Fig. 5.
This is the first case of a nucleus in the region of $A=47$ to $A=49$
(12 nuclei studied to date) where we did not obtain the
correct ground state spin with CD-Bonn + 3 terms Hamiltonian.

\subsection{Single-particle characteristics}

In order to better understand the underlying physics of
our NCSM results, we investigated the single-particle-like
properties of our solutions.

In a simple closed-shell nucleus, we expect the leading
configuration of the ground state solution in our m-scheme
treatment to be a single Slater determinant.  Single-particle
(or hole) excitations should be easily identified by the character of
their leading configurations - i.e. a single-particle creation
(or destruction) operator acting on the ground state
Slater determinant of the reference nucleus $^{48}Ca$.
For our odd-mass nuclei, this is the character
we seek.  That is, we take the standard phenomenological shell model
configuration of a single Slater determinant with a closed sd-shell
for the protons and a closed $f_{7/2}$ subshell for the neutrons
and look for the appropriate states which have a single nucleon
added to (or subtracted from) that Slater determinant.  We accept
states as "single-particle-like" when we find one with a leading
configuration having more than 50\% probability to be in the simple
configuration just described.  When the majority weight is
distributed over a few states, we use the centroid and we discuss
those cases in some detail below.  We were not successful  in
locating all the expected single-particle-like and single-hole-like states.
That is, those absent from our presentation below were spread 
among a large number of eigenstates.

For a closed-shell nucleus (Z,N) the single-particle energies (SPE)
for states above the Fermi surface are related to the binding energy
differences:\\

$e_p^{>} = BE(Z,N)-BE^{*}(Z+1,N)$,\label{equationa}\\
and

$e_n^{>} = BE(Z,N)-BE^{*}(Z,N+1)$.\label{equationb}\\

The SPE for sates below the Fermi surface are given by

$e_p^{<} = BE^{*}(Z-1,N)-BE(Z,N),$\label{equationc}\\
and

$e_n^{<} = BE^{*}(Z,N-1)-BE(Z,N)$.\label{equationd}\\

The BE are ground state binding energies which are taken as positive
values, and e will be negative for bound states.  $(BE^{*} = BE-E_{x})$
is the ground state binding energy minus the excitation energy of the
excited states associated with the single-particle states.

Experimental SPE's and the results of our analysis are shown in Fig. 6.
The experimental SPE's for protons and neutrons follow B.A.Brown's
analysis \cite{[BAB01]}.  To guide the eye, we draw a horizontal line
to indicate the vicinity of the Fermi surfaces for the protons and neutrons.

Fig. 6 shows that proton shell closure is established with both Hamiltonians,
the CD-Bonn and the CD-Bonn+3 terms.  The correct energy locations
are better approximated with the modified Hamiltonian.
Fig. 6 also shows that neutron subshell closure only appears
with modified Hamiltonian.  Here, the ordering is correct but
the states are considerably more spread out compared
with experiment.

Let us consider some of the details underlying the single-particle-like states.
The situation for the $1p_{3/2}$ or "$1p3$" state in the left panel of Fig. 6,
the proton single-particle state in $^{49}Sc$, with the modified Hamiltonian,
is quite interesting.  It appears that this state is mixed over several
excited states in the spectrum.  We can take the strength spread over
several states and construct a centroid for this $1p3$ state by a
weighted average over the states carrying that strength.  Here are
the relevant input ingredients.

The first excited state of $^{49}Sc$ is a ${3/2}^{-}$, as seen in
the second column of Fig. 4, with about 51\% of the occupancy
of the $1p3$ state.
Its eigenvalue is -425.151 MeV compared to a ground state of -428.365
MeV. The 18th state in the $^{49}Sc$ spectrum is also a ${3/2}^{-}$ with
28\% of the occupancy of the $1p3$ state.  Its eigenvalue is -422.803MeV.
The 24th state is also a ${3/2}^{-}$ with 21\% of the occupancy of
the $1p3$ state.  Its eigenvalue is -422.440MeV.

Thus, to a good approximation, the $1p3$ strength is spread over these
three states.  We will identify the weighted average $[0.51 \times (-428.365) +
0.28 \times (-422.803) + 0.21 \times (-422.440)] = -423.79$ as the
centroid of the single particle $1p3$ state which we
then include accordingly in the second column of the figure.

For the proton hole states with the modified Hamiltonian,
we perform a detailed search up to excitation energies
of about 14 MeV in the $^{47}K$ spectra.
It appears that the $0d_{5/2}$ single-hole state is spread among many states
with the largest observed concentration on the ${5/2}^{+}$ state at -386.17MeV
(13.36 MeV of excitation energy).
Here, we find a single $J^{\pi} = {5/2}^{+}$ state in $^{47}K$
with 30 \% $0d_{5/2}$ vacancy and we assign this state to our
$0d_{5/2}$ single-hole state.  Most of the $0d_{5/2}$ strength,
however, was not observed among
the limited number of converged eigenstates.

Let us consider the $^{49}Ca$ results with the modified Hamiltonian
in the upper right panel of Fig. 6.
The ground state is approximately a pure
$[(1p_{3/2})^1(0f_{7/2})^8]$ configuration.
We note that the spacing for the subshell closure is in good agreement
with experiment while there is a shift of a couple MeV towards more binding
in the model as previously indicated in Fig. 1.
A nearly pure $1p_{1/2}$ single-particle state is obtained at
5.235 MeV excitation energy and an extra low-lying ${7/2}^{-}$ appears with 2p-1h
character (see Fig. 2).  Our lowest-lying ${5/2}^{-}$ consists of 2p-1h character relative to
subshell closure.

We contrast the modified Hamiltonian's results for the $^{49}Ca$
ground state with those obtained using the $\it{ab-initio}$ CD-Bonn where
$[(1p_{3/2})^{4}(1p_{1/2})^{2}(0f_{7/2})^{3}]_{1/2^{-}}$ is the
dominant configuration reflecting again the inadequacies
of the neutron single-particle properties.





\subsection{Monopole matrix elements V(ab;T)}

\noindent
The monopole matrix element  is defined by an angular momentum
average of coupled doubly-reduced two-body matrix elements:
\begin{equation}\label{monopole}
V(ab;T) =
\frac{\sum_{J} (2J+1) V(abab:JT)}
{\sum_{J} (2J+1)} \; .
\end{equation}
For our NCSM Hamiltonians the "$V$" appearing in Eqn. \ref{monopole}
signifies the full 2-body intrinsic-coordinate Hamiltonian,
$T_{rel} + V_{eff}$, except that  we omit the Coulomb interaction
from this analysis.

We examine the monopole character of our initial CD-Bonn
Hamiltonian and we note some similarities and differences
from the GXPF1 interaction\cite{[HOB04]} as shown in Fig. 7.
The see-saw shapes of the two Hamiltonians in Fig. 7 are similar
but our Hamiltonian is shifted towards less attraction.
Given the many differences between the respective theoretical starting points,
the $\it{ab-initio} H_{eff}$ for the NCSM  and the G-matrix for GXPF1
the different bare NN interactions, etc.,  the similarities
observed in Fig. 7 are remarkable.
In order to summarize  a comparison of the underlying
theoretical interactions, we list in Table 1a simplified
overview of their differences and similarities.

For a more detailed comparison of the interactions, we present
the fp-shell matrix elements applicable to the present investigation
in Tables 2 and 3.  For convenience in finding the major differences, we
present two columns of key differences in the matrix elements:
"diff1" represents the difference between our $ab-initio ~H_{eff}$ and
our modified $H_{eff}$; and,  "diff2" represents the difference between our
modified $H_{eff}$ and the GXPF1 interaction.  While diff1 shows magnitudes
that only occasionally exceed 1 MeV, diff2 shows magnitudes approaching 2.6 MeV.
This type of comparison suggests that our solution for the modified Hamiltonian
remains closer to our initial $H_{eff}$ derived from CD-Bonn, than it is
to the fitted Hamiltonian, GXPF1.

Fig. 8 presents a similar comparison of the monopole character in the
fp-shell of the two phenomenological Hamiltonians, GXPF1 and
CD-Bonn + 3 terms. Overall, the changes in the monopole character
due to the addition of the phenomenological terms to our $H_{eff}$
of Fig. 7 appear somewhat larger for the T=1
monopole than for T=0. The effect of "+3 terms" is to increase
the T=0 and T=1 splitting of six of the monopoles while the 4 remaining
T=0 and T=1 monopole splittings are reduced.

In order to better visualize the similarities of the fp-shell matrix elements,
 we present in Figs. 9 and 10 the same comparisons shown in
 Figs. 7 and 8, respectively, with an overall shift of the GXPF1 monopole
 matrix elements so that the average over all monopole matrix elements
 is the same for the two Hamiltonians.  Specifically the average shift of T=0 and
T=1 monopoles for GXPF1 in Fig. 9 is 0.899768 MeV, while for
GXPF1 in Fig. 10 it is 0.83485 MeV. It is now evident that both the
 CD-Bonn and CD-Bonn + 3 terms have monopoles with less $T=0$
 and $T=1$ splittings.

\subsection{Matrix element correlations}

We present in Figs. 11-14 the correlations between pairs of fp-shell interaction
matrix element sets.  With Fig. 11, we observe the high degree of correlation between
the 195 matrix elements of our starting Hamiltonian, CD-Bonn, and our modified
Hamiltonian, CD-Bonn + 3 terms.  This indicates that, for the most part, our Hamiltonian
is minimally modified by the addition of the phenomenological terms.  Such a high
correlation is reminiscent of the high correlations seen between GXPF1 and its
starting interaction, the G-matrix \cite{[HOB04]}.

It is then very interesting to observe in Fig. 12 the lack of correlation between
our starting Hamiltonian, CD-Bonn, and the G-matrix underlying the GXPF1 interaction.
This lack of correlation reflects the major differences in the underlying theories that are
summarized in Table 1.

Furthermore, we find minimal correlation between CD-Bonn + 3 terms and the full GXPF1 as
seen in Fig. 13.  This indicates the likely sensitivity to the starting Hamiltonians
in the fitting procedures and to the differences in the NCSM compared to a valence
shell model approach.

Finally, in order to isolate the true off-diagonal 2-body interaction effects  from those that
may lead to contributions to the single-particle energies, we eliminate some of the
matrix elements from the comparison.  In Fig. 14 we present the correlation of 
matrix elements V(abcd; JT)(A=48) between CD-Bonn+3terms and GXPF1,
where we retain only those that cannot contribute to a single-particle Hamiltonian 
in leading order. That is, we eliminate all two-body matrix elements where at least one single-particle-state (sps) of the bra equals a sps of the ket.  There are 56 remaining two-body matrix elements.  Clearly, the correlation improves.   This suggests that those eliminated matrix elements would generate single-particle properties, in leading order, different from the single-particle properties embodied in GXPF1 plus its associated single-particle energies.  Comparisons of spectra and other properties with these Hamiltonians as one proceeds further from A=48 could shed more light on their differences.

\section{Conclusions and outlook}

We have presented an initial NCSM investigation of the spectral properties of the
$A=47$ and $A=49$ nuclei that are one nucleon away from doubly-magic
$^{48}Ca$.  We have shown that the NCSM with a previously introduced modified
Hamiltonian produces spectral properties in reasonable accord with experiment.
Shell closure properties are obtained and a path has been opened for multi-shell
investigations of these nuclei within the NCSM.  We are undertaking such additional
investigations. Also, for a better understanding of various fp-shell interactions
we made a comparison between our initial and modified fp-shell matrix elements in
the harmonic oscillator basis with the GXPF1 interaction \cite{[HOB04]}
and we found notable differences.
Our initial and modified NCSM $H_{eff}$ matrix elements in the fp-shell
are strongly correlated.  However, the same matrix elements of the modified NCSM
Hamiltonian appear to lack a significant correlation with the GXPF1 matrix elements (see Fig. 13).
We found evidence (Fig. 14) suggesting that significant differences in single-particle properties may underly some of the distinctions between our $H_{eff}$ and the GXPF1 interaction.  Additional applications could reveal those distinctions in greater detail.

\section{Acknowledgements}

This work was partly performed under the auspices of
the U. S. Department of Energy by the University of California,
Lawrence Livermore National Laboratory under contract
No. W-7405-Eng-48.
This work was also supported in part by USDOE grant DE-FG-02 87ER40371,
Division of Nuclear Physics, and in part by NSF grant INT0070789.

\newpage
\begin{table}
\begin{center}
\begin{tabular}{|c|c|c|}
\hline
\hline
Hamiltonian Property & G-matrix & NCSM cluster $H_{eff}$ \\
\hline
Oscillator parameter dependence & Yes & Yes\\
\hline
Depends on the choice of P-space & Yes & Yes\\
\hline
Translationally invariant & No & Yes\\
\hline
Reguires perturbative corrections  & & \\ and raises known convergence issues
 & Yes & No \\
\hline
Starting energy dependence & Yes & No\\
\hline
Single-particle spectra dependence & Yes & No\\
\hline
A-dependence & No & Yes\\
\hline
\hline
\end{tabular}
\end{center}
\caption{Overview of the differences and similarities of the two theoretical
approaches that underlie the Hamiltonians whose matrix elements
are compared in this work.}
\label{t1}
\end{table}

\begin{table}
\begin{center}
\begin{tabular}{|cccccccccccc|}
\hline
\hline
$2j_{a}$ & $2j_{b}$ & $2j_{c}$ & $2j_{d}$ & J & T & G & GXPF1 & diff1
& CD-Bonn & CD-Bonn & diff2\\
  & & & & & & & & & & +3 terms & \\
\hline 7 & 3 & 7 & 3 & 5 & 0 & -2.1167 & -2.8504 & -0.7337 &
-1.0390 &-1.3413 & -0.3023\\
3 & 3 & 5 & 5 & 0 & 1 & -0.5243 & -1.1968 & -0.6725 & -0.6019 &
-1.1129 & -0.5109 \\
7 & 7 & 7 & 7 & 3 & 0 & -0.2309 & -0.8087 & -0.5778 & 0.5597 &
0.5555 & -0.0042 \\
7 & 5 & 7 & 5 & 6 & 0 & -2.3465 & -2.9159 & -0.5693 & -1.3743 &
-1.8599 & -0.4856\\
7 & 5 & 7 & 5 & 5 & 0 & -0.0203 & -0.5845 & -0.5642 & 0.5813 &
0.4117 & -0.1693\\
3 & 1 & 3 & 1 & 2 & 1 & -0.7965 & -0.2822 & 0.5143 & -0.0068 &
-0.4932 & -0.4864\\
7 & 7 & 5 & 5 & 0 & 1 & -1.9095 & -1.3288 & 0.5806 & -2.2586 &
-3.4709 & -1.2123\\
\hline
\hline
\end{tabular}
\end{center}
\caption{Comparison of selected two-body matrix elements V(abcd; JT) (Mev)
(A=48) for which the difference between our interaction is large.
Diff1 represents the difference between GXPF1 and G and diff2 is the
difference between CD-Bonn+3terms and CD-Bonn.}
\label{t2}
\end{table}

\begin{table}[h]
\begin{center}
\begin{tabular}{|ccccccccccc|}
\hline
\hline
  & & & & & & & CD-Bonn  & & & \\
  $2j_{a}$ & $2j_{b}$ & $2j_{c}$ & $2j_{d}$ & J & T & CD-Bonn & +3
terms & GXPF1 & diff1 & diff2\\
\hline 7 &  7 &  7 & 7 &  1 &  0 &  0.3073 & -0.1114 & -1.2334 &
0.4187 & 1.1219\\
 7 &  7 &  7 & 7 &  3 &  0 &  0.5597 & 0.5555 & -0.8087 &
0.0042 & 1.3642\\
 7 &  7 &  7 &  7 & 5 &  0 &  0.2624 & 0.2831 & -0.7531
& -0.0207 & 1.0363\\
 7 &  7 &  7 &  7 &  7 &  0 &  -1.1278 & -1.4916 &
-2.5614 & 0.3638 & 1.0698\\
 7 &  7 &  7 &  3 &  3 &  0 &  -0.3941 &
-0.5040 & -0.8461 & 0.1099 & 0.3421\\
 7 &  7 &  7 &  3 &  5 &  0
 & -0.7265 & -0.9507 & -0.4098 & 0.2242 & -0.5409\\
  7 &  7 &  7  & 5  & 1 &  0 & 2.1173 & 2.8631 & 1.8252 & -0.7457 & 1.0379\\
 7 &  7 &  7 &  5
& 3 & 0 & 0.9514 & 1.0655 & 1.0488 & -0.1141 & 0.0167\\
 7  & 7 & 7 & 5  & 5 & 0 & 0.7734 & 0.8348 & 1.2348 & -0.0614 & -0.4001\\
7 &  7  & 7 &  1 &  3 & 0  & 0.6495 & 0.8948 & 0.8534 & -0.2453 &
0.0414\\
7 & 7 & 3 & 3 &  1 &  0 & -0.2455 & -0.3852 & -0.4144 & 0.1398 &
0.0292\\
7 & 7 & 3 & 3 & 3 & 0 & -0.3005 & -0.4174 & -0.3281 & 0.1169 &
-0.0894\\
7 & 7 & 3 & 5 & 1 & 0 & -0.0853 & -0.1703 & -0.0871 & 0.0850 &
-0.0832\\
7 & 7 & 3 & 5 & 3 & 0 & 0.1932 & 0.2031 & 0.0722 & -0.0099 &
0.1308\\
7 & 7 & 3 & 1 & 1 & 0 & 0.4140 & 0.5911 & 0.3026 & -0.1771 &
0.2885\\
7 & 7 & 5 & 5 & 1 & 0 & 1.7698 & 1.7798 & 0.6255 & -0.0100 &
1.1542\\
7 & 7 & 5 & 5 & 3 & 0 & 0.5251 & 0.3335 & 0.4187 & 0.1917 &
-0.0852\\
7 & 7 & 5 & 5 & 5 & 0 & 0.0925 & -0.1388 & 0.1190 & 0.2313 &
-0.2578\\
7 & 7 & 5 & 1 & 3 & 0 & 0.0405 & -0.0388 & -0.1040 & 0.0792 &
0.0652\\
7 & 7 & 1 & 1 & 1 & 0 & 0.1906 & 0.1891 & 0.0260 & 0.0015 &
0.1630\\
7 & 3 & 7 & 3 & 2 & 0 & 0.5237 & 0.4452 & -0.5179 & 0.0784 &
0.9632\\
7 & 3 & 7 & 3 & 3 & 0 & 0.1963 & 0.1499 & -0.9660 & 0.0464 &
1.1159\\
7 & 3 & 7 & 3 & 4 & 0 & 0.8313 & 1.0290 & -0.3550 & -0.1978 &
1.3840\\
 7& 3 & 7 & 3 & 5 & 0 & -1.0390 & -1.3413 & -2.8505 & 0.3022 & 1.5092\\
 7 & 3 &  7 & 5 & 2 & 0 & -0.8737 & -1.2298 & -0.6130 & 0.3561 & -0.6168\\
7 & 3 & 7 & 5 & 3 & 0 & 0.3768 & 0.4786 & 0.2440 & -0.1018 &
0.2346\\
7 & 3 & 7 & 5 & 4 & 0 & -0.1160 & -0.1168 & 0.1874 & 0.0008 &
-0.3042\\
7 & 3 & 7 & 5 & 5 & 0 & 0.4572 & 0.6186 & 0.6478 & -0.1614 &
-0.0292\\
7 & 3 & 7 & 1 & 3 & 0 & 1.1174 & 1.3527 & 1.6188 & -0.2353 &
-0.2661\\
7 & 3 & 7 & 1 & 4 & 0 & 0.0174 & -0.1382 & 0.1639 & 0.1556 &
-0.3021\\
7 & 3 & 3 & 3 & 3 & 0 & -0.3760 & -0.4208 & -0.4140 & 0.0447 &
-0.0068\\
7 & 3 & 3 & 5 & 2 & 0 & -1.2556 & -1.4158 & -1.2209 & 0.1602 &
-0.1949\\
7 & 3 & 3 & 5 & 3 & 0 & 0.3921 & 0.3582 & 0.5563 & 0.0339 &
-0.1981\\
7 & 3 & 3 & 5 & 4 & 0 & -0.6204 & -0.5575 & -0.6824 & -0.0629 &
0.1249\\
7 & 3 & 3 & 1 & 2 & 0 & -0.3442 & -0.3473 & -0.5983 & 0.0032 &
0.2510\\
7 & 3 & 5 & 5 & 3 & 0 & 0.3158 & 0.3418 & 0.1595 & -0.0260 &
0.1823\\
7 & 3 & 5 & 5 & 5 & 0 & 0.0470 & 0.0157 & 0.0321 & 0.0312 &
-0.0164\\
7 & 3 & 5 & 1 & 2 & 0 & 1.1771 & 1.0398 & 1.0504 & 0.1372 &
-0.0105\\
7 &  3 &  5 &  1 &  3 &  0 &  0.4204 & 0.2798 & 0.6943&
0.1407 & -0.4146\\
7 &   5 &  7 &  5  & 1 &  0 &  -3.0683 & -4.0586 & -4.4003 &
0.9903 & 0.3417\\
7 & 5 & 7 &  5 &  2 &  0 &  -1.6830 & -2.1268 & -3.1243 & 0.4438 &
0.9975\\
7 & 5 & 7 & 5 & 3 &  0 &  -0.2221 & -0.3923 & -1.3469 & 0.1702 &
0.9545\\
7 & 5 & 7 & 5 & 4 & 0 & -0.8282 & -1.3239 & -2.1696 & 0.4958 &
0.8457\\
7 & 5 & 7 & 5 & 5 & 0 & 0.5813 & 0.4117 &  -0.5845 & 0.1696 &
0.9962\\
7 & 5 & 7 & 5 & 6 & 0 & -1.3743 & -1.8599 & -2.9159 & 0.4856 &
1.0560\\
7 & 5 & 7 & 1 & 3 & 0 & -0.4566 & -0.5755 & -0.4085 & 0.1189 &
-0.1670\\
7 & 5  & 7  & 1 & 4 & 0  & -0.7299 & -0.9642 & -0.3640 & 0.2343 &
-0.6002\\

\hline
\end{tabular}
\end{center}
\end{table}

\begin{table}
\begin{center}
\begin{tabular}{|ccccccccccc|}
\hline
\hline
  & & & & & & & CD-Bonn  & & & \\
  $2j_{a}$ & $2j_{b}$ & $2j_{c}$ & $2j_{d}$ & J & T & CD-Bonn & +3
terms & GXPF1 & diff1 & diff2\\
\hline 7 & 5 & 3 & 3 & 1  & 0 & 1.0540 & 1.4541 & 0.8564 & -0.4000
&
0.5977\\
7 & 5 & 3 & 3 & 3 & 0 & 0.6228 & 0.9335 & 0.6018 & -0.3108 &
0.3317\\
7 & 5 & 3 & 5 & 1 & 0 & -0.9454 & -0.9914 & -1.2221 & 0.0461 &
0.2307\\
7 & 5 & 3 & 5 & 2 & 0 & -0.9098 & -1.3983 & -0.5745 & 0.4885 &
-0.8238\\
7 & 5 & 3 & 5 & 3 & 0 & -0.5037 & -0.7272 & -0.7413 & 0.2235 &
0.0141\\
7 & 5 & 3 & 5 & 4 & 0 & -0.7302 & -0.9373 & -0.6156 & 0.2071 &
-0.3217\\
7 & 5 & 3 & 1 & 1 & 0 & -1.6911 & -2.4985 & -1.4076 & 0.8074 &
-1.0909\\
7 & 5 & 3 & 1 & 2 & 0 & -0.8349 & -1.0867 & -0.7142 & 0.2518 &
-0.3725\\
7 & 5 & 5 & 5 & 1 & 0 & -0.7288 & -1.3474 & -0.2628 & 0.6186 &
-1.0846\\
7 & 5 & 5 & 5 & 3 & 0 & 0.5634 & 0.5072 & 0.6128 & 0.0562 &
-0.1056\\
7 & 5 & 5 & 5 & 5  & 0 & 0.9161 & 1.1081 & 1.0858 & -0.1920 &
0.0223\\
7 & 5 & 5  & 1 & 2 & 0 & 0.6576 & 0.8630 & 0.5233 & -0.2054 &
0.3397\\
7 & 5 & 5 & 1 & 3 & 0 & 0.5616 & 0.7463 & 0.6016 & -0.1846 &
0.1447\\
7 & 5 & 1 & 1 & 1 & 0 & 0.0834 & 0.2459 & 0.1852 & -0.1624 &
0.0606\\
7 & 1 & 7 & 1 & 3 & 0 & -0.4452 & -0.6265 & -1.6302 & 0.1813 &
1.0037\\
7 & 1 & 7 & 1 & 4 & 0 & 0.2353 & 0.1281 & -1.0186 & 0.1072 &
1.1467\\
7 & 1 & 3 & 3 & 3 & 0 & 0.4449 & 0.5872 & 0.6159 & -0.1422 &
-0.0288\\
7 & 1 & 3 & 5 & 3 & 0 & -0.3159 & -0.5871 & -0.0340 & 0.2712 &
-0.5531\\
7 & 1 & 3 & 5 & 4 & 0 & -1.2649 & -1.4129 & -1.3073 & 0.1480 &
-0.1056\\
7 & 1 & 5 & 5 & 3  & 0 & -0.2768 & -0.4766 & -0.2518 & 0.1998 &
-0.2248\\
7 & 1 & 5 & 1 & 3 & 0 & 0.2871 & 0.3689 & 0.4328 & -0.0818 &
-0.0639\\
3 & 3 & 3 & 3 & 1 & 0 & 0.0709 & -0.3114 & -0.6060 & 0.3822 &
0.2947\\
3 & 3 & 3 & 3 & 3 & 0 & -0.9009 & -1.0857 & -2.1991 & 0.1848 &
1.1134\\
3 & 3 & 3 & 5 & 1 & 0 & -0.1096 & -0.1431 & 0.2280 & 0.0335 &
-0.3711\\
3 & 3 & 3 & 5 & 3 & 0 & 0.1404 & 0.0837 & 0.2187 & 0.0567 &
-0.1350\\
3 & 3 & 3 & 1 & 1 & 0 & 1.7185 & 2.1121 & 1.7350 & -0.3936 &
0.3771\\
3 & 3 & 5 & 5 & 1 & 0 & 0.1123 & 0.1144 & 0.0464 & -0.0021 &
0.0680\\
3 & 3 & 5 & 5 & 3 & 0 & -0.1931 & -0.3294 & -0.0525 & 0.1363 &
-0.2769\\
3 & 3 & 5 & 1 & 3 & 0 & 0.0218 & 0.0295 & 0.1105 & -0.0077 &
-0.0810\\
3 &  3 &  1  &  1 &  1  &  0 &  0.8627 & 0.9239 &  0.7374 &
-0.0613 & 0.1866\\
3 &  5 &  3 &  5 &  1 &  0 &  -1.1592 & -1.2747 & -2.6191 & 0.1156
& 1.3444\\
3 & 5 &  3 & 5 &  2 &  0 &  -0.2345 & -0.2183 & -1.4517 & -0.0161
& 1.2333\\
3 & 5 & 3 & 5 &  3 &  0 & 0.3962 & 0.5057 & -0.5629 & -0.1095 &
1.0686\\
3 & 5 & 3 & 5 & 4 & 0  & 0.1063 & -0.0482 & -1.0455 & 0.1545 &
0.9973\\
3 & 5 & 3 & 1 & 1 & 0 & -0.3476 & -0.3027 & -0.9540 & -0.0450 &
0.6513\\
3 & 5 & 3 & 1 & 2 & 0 & -0.3270 & -0.3718 & -0.4693 & 0.0448 &
0.0976\\
3 & 5 & 5 & 5 & 1 & 0 & 0.3981 & 0.5114 & 0.4583 & -0.1134 &
0.0532\\
3 & 5 & 5 & 5 & 3 & 0 & 0.3474 & 0.5234 & 0.3074 & -0.1760 &
0.2160\\

\hline
\end{tabular}
\end{center}
\end{table}

\newpage
\begin{table}
\begin{center}
\begin{tabular}{|ccccccccccc|}
\hline
\hline
  & & & & & & & CD-Bonn  & & & \\
  $2j_{a}$ & $2j_{b}$ & $2j_{c}$ & $2j_{d}$ & J & T & CD-Bonn & +3
terms & GXPF1 & diff1 & diff2\\
\hline 3 & 5 & 5 & 1 & 2 & 0 & 0.4842 & 0.7654 & 0.3401 & -0.2812
&
0.4253\\
3 & 5 & 5 & 1 & 3 & 0 & 0.8234 & 0.9535 & 0.9752 & -0.1302 &
-0.0217\\
3 & 5 & 1 & 1 & 1 & 0 & 0.5210 & 0.5287 & 0.7817 & -0.0078 &
-0.2530\\
3 & 1 & 3 & 1 & 1 & 0 & -1.4397 & -1.9339 & -2.4084 & 0.4942 &
 0.4744\\
3 & 1 & 3 & 1 & 2 & 0 & -1.3841 & -1.8107 & -2.2214 & 0.4266 &
0.4107\\
3 & 1 & 5 & 5 & 1 & 0 & 0.1552 & 0.1834 & -0.0324 & -0.0282 &
0.2158\\
3 & 1 & 5 & 1 & 2 & 0 & 0.2759 & 0.2561 & 0.6629 & 0.0197 &
-0.4068\\
3 & 1 & 1 & 1 & 1 & 0 & 0.4259 & 0.2031 & 0.8157 & 0.2228 &
-0.6126\\
5 & 5 & 5 & 5 & 1  & 0  & 0.5809 & 0.5578 & -0.8215 & 0.0231 &
1.3793\\
5 & 5 & 5 & 5 & 3 & 0 & 0.5560 & 0.7603 & -0.5379 & -0.2042 &
1.2982\\
5 & 5 & 5 & 5 & 5 & 0 & -0.6845 & -0.8166 & -2.1920 & 0.1321 &
1.3754\\
5 & 5 & 5 & 1 & 3 & 0 & -0.4343 & -0.5141 & -0.6030 & 0.0798 &
0.0889\\
5 & 5 & 1 & 1 & 1 & 0  & -0.1464 & -0.2172 & -0.3037 & 0.0708 &
0.0864\\
5 & 1 & 5 & 1 & 2 & 0 & 0.6529 & 0.8651 & -0.3049 & -0.2122 &
1.1700\\
5 & 1 & 5 & 1 & 3  & 0 & -0.3858 & -0.4207 & -1.3472 & 0.0349 &
0.9265\\
1 & 1 & 1 & 1 & 1 & 0 & -0.0865 & -0.0553 & -1.1943 & -0.0312 &
1.1390\\
7 & 7 & 7 & 7 & 0 & 1 & -0.3350 & -2.0599 & -2.3427 & 1.7249 &
0.2829\\
7 & 7 & 7 & 7 & 2 & 1 & 0.1443 & -0.0627 & -0.8985 & 0.2070 &
0.8358\\
7 & 7 & 7 & 7 & 4 & 1 & 0.5554 & 0.5520 & -0.1245 & 0.0035 &
0.6765\\
7 & 7 & 7 & 7 & 6 & 1 & 0.7456 & 0.7885 & 0.2674 & -0.0429 &
0.5211\\
7 & 7 & 7 & 3 & 2 & 1 & -0.3418 & -0.6840  & -0.4957 & 0.3422 &
-0.1883\\
7 & 7 & 7 & 3 & 4 & 1 & -0.1909 & -0.3059 & -0.2852 & 0.1150 &
-0.0207\\
7 & 7 & 7 & 5 & 2 & 1 & -0.0952 & -0.2651 & 0.2082 & 0.1699 &
-0.4733\\
7 & 7 & 7 & 5 & 4 & 1 & -0.3340 & -0.6119 & -0.4803 & 0.2779 &
-0.1316\\
7 & 7 & 7 & 5 & 6 & 1 & -0.5566 & -0.8969 & -0.5421 & 0.3402 &
-0.3547\\
7 & 7 & 7 & 1 & 4 & 1 & -0.2668 & -0.3767 & -0.2014 & 0.1099 &
-0.1753\\
7 & 7 & 3 & 3 & 0 & 1 & -0.5835 & -1.2136 & -0.6892 & 0.6301 &
-0.5244\\
7 & 7 & 3 & 3 & 2 & 1 & -0.2066 & -0.3701 & -0.1942 & 0.1635 &
-0.1759\\
7 & 7 & 3 & 5 & 2 & 1 & -0.2271 & -0.2046 & -0.1657 & -0.0225 &
-0.0389\\
7 &   7  &  3  & 5  &  4  & 1 &  -0.2018 & -0.3594 &
-0.2137 & 0.1575 & -0.1457\\
7  & 7 &  3 &  1  & 2  & 1  & -0.1860 & -0.3622 & -0.0353 & 0.1761
& -0.3269\\
7 & 7 & 5 &  5 &  0 &  1  & -2.2587 & -3.4709 & -1.3289 & 1.2123 &
-2.1420\\
7 & 7 & 5 & 5 & 2 &  1  & -0.4434 & -1.0308 & -0.1958 & 0.5874 &
-0.8350\\
7 & 7 & 5 & 5 & 4 & 1 & -0.2330 & -0.5665 & -0.0318  & 0.3335 &
-0.5347\\
7 & 7 & 5 & 1 & 2 & 1 & -0.3465 & -0.5240 & -0.1244 & 0.1776 &
-0.3996\\
7 & 7 & 1 & 1 & 0 & 1 & -0.5140 & -0.9231 & -0.3651 & 0.4091 &
-0.5580\\
7 & 3 & 7 & 3 & 2 & 1 & 0.1202 & -0.3894 & -0.5842 & 0.5096 &
0.1948\\
7 & 3 & 7 & 3 & 3 & 1 & 0.6896 & 0.9356 & 0.1500 & -0.2460 &
0.7857\\
\hline
\end{tabular}
\end{center}
\end{table}

\newpage
\begin{table}
\begin{center}
\begin{tabular}{|ccccccccccc|}
\hline
\hline
  & & & & & & & CD-Bonn  & & & \\
  $2j_{a}$ & $2j_{b}$ & $2j_{c}$ & $2j_{d}$ & J & T & CD-Bonn & +3
terms & GXPF1 & diff1 & diff2\\
\hline 7 & 3 & 7 & 3 & 4 & 1 & 0.6401 & 0.6043 & -0.1343 & 0.0357
&
0.7387\\
7 & 3 & 7 & 3 & 5 & 1 & 0.7681 & 1.4989 & 0.5686 & -0.7307 &
0.9303\\
7 & 3 & 7 & 5 & 2 & 1 & -0.0119 & -0.1848 & 0.0921 & 0.1730 &
-0.2769\\
7 & 3 & 7 & 5 & 3 & 1 & -0.0815 & -0.3396 & -0.5025 & 0.2581 &
0.1629\\
7 & 3 & 7 & 5 & 4 & 1 & -0.1517 & -0.2794 & -0.2388 & 0.1277 &
-0.0405\\
7 & 3 & 7 & 5 & 5 & 1 & 0.0043 & -0.5487 & -0.4621 & 0.5531 &
-0.0866\\
7 & 3 & 7 & 1 & 3 & 1 & 0.0600 & -0.1725 & -0.1007 & 0.2325 &
-0.0718\\
7 & 3 & 7 & 1 & 4 & 1 & -0.3353 & -0.6527 & -0.3219 & 0.3173 &
-0.3307\\
7 & 3 & 3 & 3 & 2 & 1 & -0.2402 & -0.3588 & -0.3591 & 0.1185 &
0.0004\\
7 & 3 & 3 & 5 & 2 & 1 & -0.4330 & -0.5878 & -0.5223  & 0.1548 &
-0.0655\\
7 & 3 & 3 & 5 & 3 & 1 & 0.0012 & 0.1176 & 0.1764 & -0.1164 &
-0.0587\\
7 & 3 & 3 & 5 & 4 & 1 & -0.5820 & -0.8194 & -0.4367 & 0.2374 &
-0.3827\\
7 & 3 & 3 & 1 & 2 & 1 & -0.2734 & -0.3242 & -0.4095 & 0.0508 &
0.0852\\
7 & 3 & 5 & 5 & 2 & 1 & -0.5131 & -0.6285 & 0.0845 & 0.1154 &
-0.7131\\
7 & 3  & 5 & 5 & 4 & 1 & -0.1958 & -0.2548 & -0.2062 & 0.0590 &
-0.0486\\
7 & 3 & 5 & 1 &  2 & 1 & -0.8563 & -1.3191 & -0.7715 & 0.4628 &
-0.5476\\
7 & 3 & 5 & 1 & 3 & 1 & -0.0252 & -0.0957 & -0.1743 & 0.0705 &
0.0786\\
7 & 5 & 7 & 5 & 1 & 1 & 0.5082 & 2.4886 & -0.0854 & -1.9804 &
2.5740\\
7 & 5 & 7 & 5 & 2 & 1 & 0.7226 & 0.6164 & -0.1681 & 0.1062 &
0.7845\\
7 & 5 & 7 & 5 & 3 & 1 & 0.6587 & 1.8189 & 0.6055  & -1.1602 &
1.2134\\
7 & 5 & 7 & 5 & 4 & 1 & 0.6129 & 0.4423 & 0.4576 & 0.1706 &
-0.0153\\
7 & 5 & 7 & 5 & 5 & 1 & 0.6787 & 1.6340 & 0.7141 & -0.9554 &
0.9199\\
7 & 5 & 7 & 5 & 6 & 1 & -0.3906 & -1.0421 & -0.9527 & 0.6515 &
-0.0895\\
7 & 5 & 7 & 1 & 3 & 1 & -0.0360 & 0.6988 & 0.3097 & -0.7349 &
0.3891\\
7 & 5 & 7 & 1 & 4 & 1 & -0.0990 & -0.2726 & 0.1832 & 0.1736 &
-0.4558\\
7 & 5 & 3 & 3  & 2 & 1 & -0.0618 & 0.0096 & 0.0689  & -0.0715
 & -0.0593\\
7 & 5 & 3 & 5 & 1 & 1 & -0.1021 & 0.5367 & 0.0501 & -0.6388 &
0.4866\\
7 & 5 & 3 & 5 & 2 & 1 & -0.1821  & -0.3153 & -0.4080 & 0.1332 &
0.0927\\
7 & 5 & 3 & 5 & 3 & 1 & -0.0857 & -0.1165 & -0.0257 & 0.0308 &
-0.0907\\
7 &  5 &  3 &  5 & 4  &  1 &  -0.4104 &  -0.5206 &  -0.2593
& 0.1101 & -0.2613\\
7 &  5 & 3 &  1 &  1  & 1 &  -0.0507 & -0.4156 & 0.0530 & 0.3649 &
-0.4686\\
7 & 5 & 3 & 1 &  2 &  1 &  -0.1190 & -0.3124 & -0.0147 & 0.1934 &
-0.2977\\
7 & 5 & 5 & 5 & 2  & 1 &  -0.4658 &  -0.6595 &  -0.4825 & 0.1936 &
-0.1770\\
7 & 5 & 5 & 5 & 4 & 1 & -0.3706 &  -0.5363 &  -0.2603 & 0.1657 &
-0.2760\\
7 & 5 & 5 & 1 & 2 & 1 & -0.3356 & -0.3388 &  -0.1477 & 0.0032 &
-0.1911\\
7 & 5 & 5 & 1 & 3 & 1 & 0.0577 & -0.2258 & 0.1062 & 0.2835 &
-0.3320\\
7 & 1 & 7 & 1 & 3 & 1 & 0.7343 & 1.4734 & 0.4682 & -0.7391 &
1.0052\\
7 & 1 & 7 & 1 & 4 & 1 & 0.4269 & 0.3100 & -0.1294 & 0.1169 &
0.4394\\

\hline
\end{tabular}
\end{center}
\end{table}

\newpage
\begin{table}
\begin{center}
\begin{tabular}{|ccccccccccc|}
\hline
\hline
  & & & & & & & CD-Bonn  & & & \\
  $2j_{a}$ & $2j_{b}$ & $2j_{c}$ & $2j_{d}$ & J & T & CD-Bonn & +3
terms & GXPF1 & diff1 & diff2\\
\hline 7 & 1 & 3 & 5 & 3 & 1 & 0.0358 & 0.3366 & 0.3738 & -0.3009
&
-0.0372\\
7 & 1 & 3 & 5 & 4 & 1 & -0.6006 & -0.9045 & -0.5871 & 0.3039 &
-0.3174\\
7 & 1 & 5 & 5 & 4 & 1 & -0.2879 & -0.3606 & -0.2160 & 0.0727 &
-0.1446\\
7 & 1 & 5 & 1 & 3 & 1 & -0.0740 & -0.0480 & -0.1524 & -0.0260 &
0.1044\\
3 & 3 & 3 & 3 & 0 & 1 & -0.2470 & -1.4569 & -1.0727 & 1.2099 &
-0.3842\\
3 & 3 & 3 & 3 & 2 & 1 & 0.3792 & 0.1916 & -0.0852 & 0.1875 &
0.2768\\
3 & 3 & 3 & 5 & 2 & 1 & -0.0594 & -0.0203 & -0.4449 & -0.0390 &
0.4246\\
3 & 3 & 3 & 1 & 2 & 1 & -0.5458 & -0.9685 & -0.6091 & 0.4227 &
-0.3594\\
3 & 3 & 5 & 5 & 0 & 1 & -0.6019 & -1.1128 & -1.1968 & 0.5109 &
0.0839\\
3 & 3 & 5 & 5 & 2 & 1 & -0.0825 & -0.2487 & 0.0691 & 0.1662 &
-0.3177\\
3 & 3 & 5 & 1 & 2 & 1 & -0.1325 & -0.1434 & -0.1847 & 0.0109 &
0.0414\\
3 & 3 & 1 & 1 & 0 & 1 & -1.2996 & -2.0501 & -1.4342 & 0.7506 &
-0.6160\\
3 & 5 & 3 & 5 & 1 & 1 & 0.6068 & 1.5042 & 0.3155 & -0.8974 &
1.1887\\
3 & 5 & 3 & 5 & 2 & 1 & 0.8432 & 0.8865 & 0.3466 & -0.0432 &
0.5398\\
3 & 5 & 3 & 5 & 3 & 1 & 0.7102 & 1.4885 & 0.3324 & -0.7783 &
1.1561\\
3 & 5 & 3 & 5 & 4 & 1 & 0.4260 & 0.1411 & -0.2483 & 0.2848 &
0.3894\\
3 & 5 & 3 & 1 & 1 & 1 & -0.0955 & 0.5019 & -0.1034 & -0.5975 &
0.6053\\
3 & 5 & 3 & 1 & 2 & 1 & -0.1390 & -0.2215 & -0.4367 & 0.0825 &
0.2151\\
3 & 5 & 5 & 5 & 2 & 1 & -0.0040 & -0.2060 & -0.0538 & 0.2021 &
-0.1522\\
3 & 5 & 5 & 5 & 4 & 1 & -0.0532 & -0.2108 & -0.3473 & 0.1576 &
0.1365\\
3 & 5 & 5 & 1 & 2 & 1 & -0.2637 & -0.5793 & -0.3884 & 0.3156 &
-0.1909\\
3 & 5 & 5 & 1 & 3 & 1 & -0.0287 & 0.4308 & 0.0576 & -0.4595 &
0.3732\\
3 & 1 & 3 & 1 & 1 & 1 & 0.7211 & 1.5211 & -0.1531 & -0.8000 &
1.6742\\
3 & 1 & 3 & 1 & 2 & 1 & -0.0068 & -0.4932 & -0.2823 & 0.4864 &
-0.2109\\
3 & 1 & 5 & 5 & 2 & 1 & -0.2235 & -0.3536 & 0.0576 & 0.1301 &
-0.4112\\
3 & 1 & 5 & 1 & 2 & 1 & -0.2449 & -0.3398 & -0.2392 & 0.0950 &
-0.1006\\
5 & 5 & 5 & 5 & 0 & 1 & 0.3171 & -1.0579 & -1.1607 & 1.3749 &
0.1028\\
5 & 5 & 5 & 5 & 2 & 1 & 0.5129 & 0.5285 & -0.4440 & -0.0156 &
0.9724\\
5 & 5 & 5 & 5 & 4 & 1 & 0.8190 & 0.9119 & -0.1560 & -0.0929 &
1.0679\\
5 & 5 & 5 & 1 & 2 & 1 & -0.1027 & -0.3641 & -0.3082 & 0.2614
 & -0.0559\\
5 & 5 & 1 & 1 & 0 & 1 & -0.3086 & -0.7120 & -0.7775 & 0.4034 &
0.0656\\
5 & 1 & 5 & 1 & 2 & 1 & 0.5515 & 0.3998 & -0.1459 & 0.1517 &
0.5457\\
5 & 1 & 5 & 1 & 3 & 1 & 0.7763 & 1.3514 & 0.2289 & -0.5751 &
1.1224\\
1 & 1 & 1 & 1 & 0 & 1 & 0.6719 & -0.0072 & -0.4294 & 0.6791 &
0.4222\\
\hline
\hline
\end{tabular}
\end{center}
\caption{Comparison of the fp-shell two-body matrix elements V(abcd; JT) (Mev)
(A=48) employed in this work. The interaction GXPF1 is taken from Ref. \cite{[HOB04]}. 
Diff1 represents the difference between CD-Bonn+3terms and CD-Bonn
and Diff2 is the difference between CD-Bonn+3terms and GXPF1.}
\label{t3}
\end{table}

\end{document}